\documentclass[twoside,narrowdisplay]{newelsart}
\usepackage{amsfonts}
\usepackage{amsmath}
\usepackage{amssymb}
\journal{@arXiv.org}
\renewcommand{\thesubsection}{\arabic{section}.\arabic{subsection}.\hspace{-4mm}}
\newtheorem{theorem}{The theorem}
\newcommand{\theoremcaption}[1]{\bf~is~#1:\it}
\newcommand{\mathtab}{\\[6pt]}
\newtheorem{proof}{}
\newcommand{\proofcaption}[1]{\bf The theorem \ref{#1} proof: \it}
\newtheorem{postulate}{The postulate}[section]
\newcommand{\postulatecaption}[1]{~\bf is asserting #1:\rm}
\newcommand{\theoremref}[1]{in the theorem~\ref{#1}}
\newcommand{\postulateref}[1]{in the postulate~\ref{#1}}
\newcommand{\eqoref}[1]{Eq.~(\ref{#1})}
\newcommand{\eqsref}[2]{Eqs.~(\ref{#1})--(\ref{#2})}
\newcommand{\Jppbib}[4]{{\bf #1}~(#2)~#3--#4}
\newcommand{\Jbib}[3]{{\bf #1}~(#2)~#3}
\newcommand{\Bbib}[2]{#1,~#2}
\newcommand{\tensum}[4]{\displaystyle{{2^{- \left(2\, m\, +\,
1\right)}}\over{n\left(x^{\;\alpha}\right)\; Q_{c}}}\left[
\hat{#1}^{\;#3,\;\ldots\;,\;#4}\;\hat{#2}_{\;#3,\;\ldots\;,\;#4}
+\hat{#2}_{\;#3,\;\ldots\;,\;#4}\;\hat{#1}^{\;#3,\;\ldots\;,\;#4}\;
\right]}
\newcommand{\divergency}[1]{\partial^{\;#1}}
\newcommand{\divergencyc}[1]{\partial_{\;#1}}
\newcommand{\Sdiv}[1]{\lambda_{c}\;\divergency{#1}}
\newcommand{\Sdivc}[1]{\lambda_{c}\;\divergencyc{#1}}
\newcommand{\Ddiv}[1]{\lambda^{#1}_{c}\;\divergency{#1}}
\newcommand{\FactorF}[4]{k_{#4}\;=\;1\;+\;{#1}_{\;#2}{#1}^{\;#2}\;+\;\left[\Sdiv{#3}{#1}_{\;#3}\right]\;+}
\newcommand{\FactorS}[3]{\left[\left(\Ddiv{2}{#1}_{\;#2}\right)\left(\Ddiv{2}{#1}_{\;#2}\right)\right]\;+
\;\left[\Sdiv{#3}{#1}_{\;#3}\right]^2}
\newcommand{\Tdiv}[5]{\hat{#5}^{\;#3,\;\ldots\;,\;#4}_{;\;{#1}}\;=\;
\displaystyle{{1}\over {2}}\;\left(\;\Sdivc{#1}
\;\hat{#2}^{\;#3,\;\ldots\;,\; {#1},\;\ldots\;,\;#4}\right)}
\newcommand{\streamF}[6]
{{#6}^{#1}\;=\;\displaystyle{{2^{- \left(2\, m\, +\,
1\right)}}\over{ n \left( x^{\;\alpha}\right)\;
Q_{c}}}\;{\sum\limits_{#1\;=\;{#4}}^{\;#5}\;\left(
\hat{#2}^{\;#4,\;\ldots\;,\;#5}_{;\;#1}\;\hat{#3}_{\;#4,\;\ldots\;,\;
{#1},\;\ldots\;,\;#5}\right) \;+}}
\newcommand{\streamS}[5]
{\displaystyle{{2^{- \left(2\, m\, +\, 1\right)}}\over {n \left(
x^{\;\alpha}\right)\;
Q_{c}}}\;{\sum\limits_{#1\;=\;{#4}}^{\;#5}\;\left(\hat{#3}_{\;#4,\;\ldots\;,\;{#1},\;\ldots\;,\;#5}
\;\hat{#2}^{\;#4,\;\ldots\;,\;#5}_{;\;#1}\right)}}
\newcommand{\tensor}[9]
{\hat{#1}^{\; #2,\; #3,\;\ldots\;,\; #4,\; #5}=\; {\Lambda}^{\;
#2}_{\; #6}\;{\Lambda}^{\; #3}_{\; #7}\;\ldots\;{\Lambda}^{\;
#4}_{\; #8}\;{\Lambda}^{\; #5}_{\; #9}\;
\hat{#1}^{\;#6,\;#7,\;\ldots\;,\;#8,\;#9}}
\newcommand{\PsidevF}[6]
{{#1}_{\;#2}=\displaystyle{{2^{- \left(2\; m\; +\; 1\right)}}\over
{n \left( x^{\;\alpha}\right)\; Q_{c}}}
\;{\sum\limits_{\nu\;=\;{#5}}^{\;#6}}
\hat{#3}^{\;#5,\;\ldots\;,\;#6}_{;\;\nu}\;\hat{#4}_{\;#5,\;\ldots\;,\;#6;\;\nu}\;+}
\newcommand{\PsidevS}[4]{
\displaystyle{{2^{- \left(2\, m\, +\, 1\right)}}\over{n \left(
x^{\;\alpha}\right)\;Q_{c}}} \;{\sum\limits_{\nu\;=\;{#3}}^{\;#4}
\hat{#1}^{\;#3,\;\ldots\;,\;#4}_{;\;\nu}\;\hat{#2}_{\;#3,\;\ldots\;,\;#4;\;\nu}}}
\newcommand{\IBdifF}[5]{{#1}\;=\;{#2}^{#4}{#3}_{#4}\;+\;\Sdiv{#5}{#2}_{#5}\;+}
\newcommand{\IBdifS}[5]{\left(\Sdiv{#3}{#1}_{#3}\right)\left(\Sdiv{#4}{#2}_{#4}\right)\;
+\mathtab\left(\Ddiv{2}{#1}_{#5}\right)\left(\Ddiv{2}{#2}^{#5}\right)}
\newcommand{\JHFdif}[5]{{#1}\;=\;{#2}^{\;#4,\;#5}{#3}_{\;#4,\;#5}\;+\;\tilde{#2}^{\;#4,\;#5}\tilde{{#3}}_{\;#4,\;#5}\;
+\;\displaystyle{{1}\over
{2}}\;\left[{#2}^{\;#4,\;#5}\tilde{{#3}}_{\;#4,\;#5}\;+\tilde{#2}^{\;#4,\;#5}{{#3}}_{\;#4,\;#5}\right]+\;}
\newcommand{\JHSdif}[4]{\left(\Sdiv{#3}{#1}_{\;#3,\;#4}\right)\left(\Sdiv{#3}{#2}_{\;#3}^{\;#4}\right)\;+
\;\left(\Sdiv{#3}\tilde{#1}_{\;#3,\;#4}\right)\left(\Sdiv{#3}\tilde{#2}_{\;#3}^{\;#4}\right)\;+}
\newcommand{\JHTdif}[4]
{\displaystyle{{1}\over
{2}}\;\left(\Sdiv{#3}\tilde{#1}_{\;#3,\;#4}\right)\left(\Sdiv{#3}{#2}_{\;#3}^{#4}\right)\;+\;
\displaystyle{{1}\over
{2}}\left(\Sdiv{#3}{#1}_{\;#3,\;#4}\right)\left(\Sdiv{#3}\tilde{#2}_{\;#3}^{#4}\right)}
\newcommand{\IBCFdif}[4]{{#1}\;=\;\displaystyle{{1}\over
{2}}\;\left[{#2}^{#4}{\;#3}_{\;#4}\;+\;{#3}^{\;#4}{#2}_{\;#4}\right]\;+}
\newcommand{\IBCSdif}[4]{
\displaystyle{{1}\over
{2}}\;\left[\left(\Sdiv{#3}{#1}_{\;#3}\right)\left(\Sdiv{#4}{#2}_{\;#4}\right)\;
+\;\left(\Sdiv{#4}{#2}_{\;#4}\right)\left(\Sdiv{#3}{#1}_{\;#3}\right)\right]\;+}
\newcommand{\IBCTdif}[3]{\displaystyle{{1}\over
{2}}\;\left[\left(\Ddiv{2}{#1}_{\;#3}\right)\left(\Ddiv{2}{#2}^{\;#3}\right)\;+
\;\left(\Ddiv{2}{#2}_{#3}\right)\left(\Ddiv{2}{#1}^{\;#3}\right)\right]}
\newcommand{\IBJHFdif}[7]
{{#1}\;=\;\displaystyle{k_{#7}}\left[{#2}_{\;#5}\;{#4}^{\;#5,\;#6}\;{#3}_{\;#6}\;
+\;{#2}_{\;#5}\;\tilde{#4}^{\;#5,\;#6}\;{#3}_{\;#6}\right]\;+}
\newcommand{\IBJHSdif}[5]
{\displaystyle{{k_{#1}}\over {2}}\;
\left[\left(\Sdiv{#4}{#2}_{\;#4,\;#5}\right){#3}^{\;#5}\;
+\;{#3}^{\;#5}\;\left(\Sdiv{#4}{#2}_{\;#4,\;#5}\right)\right]\;+\mathtab
\displaystyle{{k_{#1}}\over {2}}\;
\left[\left(\Sdiv{#4}{#2}_{\;#5,\;#4}\right){#3}^{\;#5}\;
+\;{#3}^{\;#5}\;\left(\Sdiv{#4}{#2}_{\;#5,\;#4}\right)\right]\;+}
\newcommand{\IBJHTdif}[5]
{\displaystyle{k_{#5}}\;\left(\Ddiv{2}{#1}^{\;#3}\right){#2}_{\;#3,\;#4}\left(\Ddiv{2}{#1}^{\;#4}\right)\;
+\mathtab
\displaystyle{k_{#5}}\;\left(\Ddiv{2}{#1}^{\;#3}\right)\;\tilde{{#2}}_{\;#3,\;#4}\left(\Ddiv{2}{#1}^{\;#4}\right)\;+}
\newcommand{\IBJHFodif}[6]
{\displaystyle{{k_{#6}}\over
{2}}\;\left(\Sdiv{#4}\tilde{#2}_{\;#4,\;#3}\right)\left(\Sdiv{#4}{#2}_{\;#4}^{\;#3}\right)\;+\mathtab
\displaystyle{{k_{#6}}\over
{2}}\;\left(\Sdiv{#4}\tilde{#2}_{\;#3,\;#4}\right)\left(\Sdiv{#4}{#2}^{\;#3}_{\;#4}\right)}
\newcommand{\JHCAlldif}[9]
{{#1}\;=\;
\displaystyle{{1}\over{2}}\left[{#2}^{\;#4,\;#5}{#3}_{\;#4,\;#5}\;+\;{#3}^{\;#4,\;#5}{#2}_{\;#4,\;#5}\right]\;+\mathtab
\displaystyle{{1}\over{2}}\left[\tilde{#2}^{\;#6,\;#7}\tilde{{#3}}_{\;#6,\;#7}\;+
\;\tilde{#3}^{\;#6,\;#7}\tilde{{#2}}_{\;#6,\;#7}\right]\;+\mathtab
\displaystyle{{1}\over{2}}\left[\tilde{#2}^{\;#4,\;#5}{{#3}}_{\;#4,\;#5}\;+
\;{#3}^{\;#4,\;#5}\tilde{{#2}}_{\;#4,\;#5}\right]\;+\mathtab
\displaystyle{{1}\over{2}}\left[{#2}^{\;#8,\;#9}\tilde{{#3}}_{\;#8,\;#9}\;+
\;\tilde{#3}^{\;#8,\;#9}{{#2}}_{\;#8,\;#9}\right]}
\begin{document}
\begin{frontmatter}
\title{The Description of Information in 4-Dimensional
Pseudo-Euclidean Information Space}
\author[PSATI]{O.I.~Shro\corauthref{cor}}
\corauth[cor]{Corresponding author.}
\ead{oshro@psati.ru}
\address[PSATI]{Volga State Academy of Telecommunications and Informatics,
Department of Information Systems and Technologies, Chair of The
Software and Control of Technical Systems, 23 L. Tolstogo
str., Samara, 443011, Russian} %\maketitle

\begin{abstract}
This article is presented new method of description information
systems in abstract 4-dimensional pseudo-Euclidean information space
(4-DPIES) with using special relativity (SR) methods. This purpose
core postulates of existence 4-DPIES are formulated. The theorem
setting existence criteria of the invariant velocity of the
information transference is formulated and proved. One more theorem
allowed relating discrete parameters of information and continuous
space-time treating and also row of supplementary theorems is
formulated and proved. For description of dynamics and interaction
of information, in article is introduced general parameter of
information - generalized information emotion (GIE), reminding
simultaneously on properties the mass and the charge. At performing
calculation of information observable parameters in the information
space is introduced continual integration methods of Feynman. The
applying idea about existence of GIE as measures of the information
inertness and the interaction carrier, and using continual
integration methods of Feynman can be calculated probability of
information process in 4-DPIES. In this frame presented approach has
allowed considering information systems when interest is presented
with information processes, their related with concrete definition
without necessity. The relation between 4-DPIES and real systems
parameters is set at modelling of matching between observable
processes and real phenomena from information interpretation.
\end{abstract}
\end{frontmatter}

\section{Introduction}
The consideration renderings of the information, the description of
information dynamics and coursing processes at participation of the
information, despite lacking a precise and unequivocal determination
of the information are the important problems
\cite{Neg84,Tur84,Cod70,Aul05,HoO05,HoS05,She56}. Importance of the
problems related with a treating of the information is gathered in
wide use of renderings about information processes in systems. In
many frames as such information processes real physical and chemical
processes act. However at a studying of information interchanges
between systems there is a problem of studying of the most
information process, instead of physical and chemical processes
laying in its bottom is more often. The instances of such problems
on studying of information systems can acting problems of
generation, transfer and perception of information
\cite{Neg84,Tur84,Cod70,Aul05,HoO05,HoS05,She56,She48,She49,She50,She51,She49MP,
Kol91,Haz98,AsZ07,JoV07,HiM07,Mas07,Kad94,Kad03}, for example;
problems of relational databases modeling \cite{ChR07,LeB07};
problems of a microbiology and biogenetic informatics
\cite{ArL96,MaC97,HaC98,Kan02,BaV04,Kan04,Koc04,MiB03,Gim06};
studying of interaction of the information on logic and social
systems development
\cite{CoW05,Ben05,Kur06,Zad65,Zad68,LeZ69,Zad75I,Zad75II,Zad05,
GaP04,Qiu05,PeR06,SeC07,HuC07,DeC07}, in particular in particular
closely related with objects cultural science
\cite{BoR96,HeB01BCFGM,RiB01,HeB01,BoR05,MoC07} and linguistics
\cite{Tat89,Koz91,LoB05,FaB06,Bur06,Per07,Zie07,Mik07}.

In present time the core used approach is the treating of the
information as measures entropy of source in which bottom the ideas
tendered Sannon lay \cite{She48} the got developments in follow-on
articles \cite{She49,She50,She51,She49MP,
Kol91,Haz98,AsZ07,JoV07,HiM07,Mas07}. The bases of these approach
attempts of information inclusion in treating physical systems,
viewed in articles are undertaken \cite{Aul05,HoO05,HoS05,Haz98}. In
B.B.Kadomtseva's article \cite{Kad94} is presented the approach
based on the analysis of irreversibility of the physical processes
observed in system. Thus observed quantum-mechanical effects are
identified with information processes \cite{Kad94,Kad03}.

The consideration of the information based on studying of properties
of physical systems, there is a row of essential difficulties.
First, the account of time characteristics of the information at not
relativistic studying of physical, chemical, biological and social
systems with which the information and methods of an information
exchange is related provided that viewed systems are relativistic,
so for example, the electromagnetic wave by means of which the
information extends \cite{Aul05,HoO05,HoS05,Kad94,Kad03}. Secondly,
difficulties in studying the information object and subject
relation, as rule is attitudes object-subject allocation
complexities \cite{Neg84,Tur84}. Thirdly, complexities of the
account of logical interference of the information which can be not
always shown to properties of physical, chemical, biological and
social systems.

In particular, the third complexity has led to a development of the
device of discrete mathematics, boolean logic
\cite{CoW05,Ben05,Kur06}, and also L.A~Zadeh's articles tendered in
series ideas fuzzy sets, fuzzy logic \cite{Zad65,Zad68,LeZ69} and
linguistic variables \cite{Zad75I,Zad75II}. On the basis of
application of the device of logic and fuzzy logic attempts of
thinking processes modelling in particular are undertaken
\cite{Zad65,Zad68,LeZ69,Zad75I,Zad75II,Zad05,GaP04,Qiu05,PeR06,SeC07,HuC07,DeC07}.

In present articles the method of the information and information
systems modelling is offered in some abstract information space, in
this frame 4-dimensional pseudo-Euclidean information space
(4-DPIES).

Introducing 4-DPIES can be proved that at description of information
relate with real physical (chemical, biological, social, etc.)
systems. Some processes and phenomena in this system is possible
compare with three-dimensional coordinates, the time in real system
identically to compare due course in information system.

The analogous situation takes place in case of physical systems
relativistic description on the basis of fundamental symmetry
properties. For example are mesons which relating two-body systems:
quark and antiquark. In the relativistic approach generators of
Poincare transformations group contact algebra physical observable
\cite{Dir49,Pol89}. The such articles are theoretical articles about
inclusion of interaction operator in algebra observable in
relativistic Hamiltonian dynamics (RHD)
\cite{Pol89,KeP91,Lev93,Lev95} and articles, devoted to studying of
electroweak mesons properties in instant form RHD
\cite{KrT93,BaK96,Kru97,BaK00,KrS01}.

The information existing on the one hand as some objective reality,
as existing physical (chemical, biological) system to which is
related, in particular the electromagnetic wave by means of which
transmission of the information is carried out. On the other hand
generation and perception of the information depends on features of
the subject of generation or perception of the information, on what
in the articles specified A.N.~Kolmogorov \cite{Kol91}.

The introduction of information space comparison of real properties
and processes of considered system to coordinate system, used is
made at the description of behaviour of the information in
information space. For relate of information space with physical
system equivalence of time in physical system in due course in an
information system is established, and for three-dimensional
coordinates in information space relate with the phenomena or
processes in real system which can be interpreted from the point of
view of the description of the information as changes of coordinates
is established.

This article consists of eight parts (paragraphs), in introduction
is given brief field review where research methods on the basis of
processes in information interpretation are widely applied
consideration.

In the second part, the question of criterion existence is
considered for introduction of invariant velocity of information
transference in information space and the basic postulates for
information space. Existence of such velocity allows entering on
4-DPEIS of Poincare group transformation at transition from one
frame of reference into another where as velocity of light in vacuum
is used invariant velocity of information transference or {\it the
light information velocity (LIV)}

In the third scientific article, general provisions of the
description of kinematics of the information in 4-DPEIS are
considered, and the criterion allowing are established to relate the
discrete parameters of observable information displacement and the
entered continuous information space.

In the fourth part, the description of information processes on the
basis of representation about existence {\it the generalized
information emotion} \cite{Kad94} is entered by formal forms.
Entered thus {\it the generalized information emotion} is
simultaneously inertial measure of information system (i.e. mass of
system by physical analogy) and a measure of interaction (charge of
system by physical analogy) on the basis of this introduction is
possible to involve for the description of information processes in
4-dimensional space methods of Maxwell's electrodynamics
\cite{HoG06,Ple75,ScS77,AsC06}. The basic equations relating the
general information parameters and entered in article {\it
generalized information emotion} are postulated. All the processes
related to generation and perception of the information, can be
represented as interaction of one information on another due to
interaction against each other {\it the generalized information
emotions}, concerning to different information.

In the fifth part, the question on structure {\it the generalized
information emotion} is considered. For a basis division of the
information into the text, implied sense and a context is taken used
in the theory of linguistics
\cite{Tat89,Koz91,LoB05,FaB06,Bur06,Per07,Zie07,Mik07}. Thus for the
description of the information in the form of the text is entered
tensor of text any rank, and for the description of the subject of
information perception is entered corresponding tensor of the
information perception. The relating between the generalized
information emotion and contributions from the text, implied sense
and context of the existing information is established. For
corresponding contributions relating between text and context are
resulted by the account heterogeneity of continuous space-time and
changes of corresponding contributions. From representation that
implied sense is functionally dependent on the text, context and
point of continuous space-time in which there is perception of the
text, relate with implied sense and corresponding contributions from
the text, context and point of continuous space-time is established.
Introduction for the parameters of the information tensors of text,
text perception, and also the generalized information emotion
inherently is close to definition linguistic variable by means of
which now are modelled information interaction of biogenetic
\cite{ArL96,MaC97,HaC98,Kan02,BaV04,Kan04,Koc04,MiB03,Gim06} and
processes responsible for system of logic and consciousness
\cite{CoW05,Ben05,Kur06,Zad65,Zad68,LeZ69,Zad75I,Zad75II,Zad05,
GaP04,Qiu05,PeR06,SeC07,HuC07,DeC07}, and also cultures science and
linguistics \cite{BoR96,HeB01BCFGM,RiB01,HeB01,BoR05,MoC07,
Tat89,Koz91,LoB05,FaB06,Bur06,Per07,Zie07,Mik07}.

In the sixth part, being based on continual integration methods,
offered Feynman \cite{FeV00} for calculation of probabilities of
physical processes depending on observable and parameters of studied
system, the question on probability of information processes course
is considered \cite{Kad94} . At modelling information processes
besides their likelihood character ambiguity modelling tensors of
text and text perception, and also senselessness, within the limits
of information problems, consideration of the free text without
taking into account perception is considered. On the basis of the
aforesaid the problem about free transference of the information
(i.e. when transference and perception of the information are not
interaction with other information), with calculation in general
view of process probability and probability density is considered.
The classification of possible problems is spent.

In the seventh part, using earlier considered for a frame of free
information transference technique, frame in point on transference
and interference between two and more information. The general
problems which can be solved are considered, applying given methods.

In the eighth, final, parts generalization of the basic results are
resulted and drawn conclusions on application of the given approach
for the decision of the general-theoretical and practical problems
related information processes.

\section{The general postulates of the 4-DPEIS description }
\label{SecPostulate} The information system in the article is
understood as the information and space in which this information
extends. At the description of information systems is necessary to
set dimension of space, to enter frame of references and to define
group of transformations of one frame of reference in another.

The time appears in modern methods of the information systems
description as free parameter of system. It leads to that the
account of system dynamics is represented complex and up to the end
not resolved problem containing set of parameters, demanding a
substantiation of introduction and criteria of admissible values
selection
\cite{Neg84,Tur84,Aul05,HoO05,HoS05,She48,She49,She50,She51,She49MP,
Kol91,Haz98,AsZ07,JoV07,HiM07,Mas07}.

The article is offered to construct 4-DPEIS where time is considered
not as free parameter, and the coordinate of the given space
directly compared in due course in real physical system to which the
entered information space is related. Such space is 4-DPEIS on which
and representation of Poincare transformations group, for the
description of transition from one frame of reference the
4-dimensional coordinates grid (the coordinates grid is used time
and three spatial coordinates) is entered into another.

The general problem of such space introduction is existence or
absence of the invariant velocity. By consideration of physical
systems by such invariant velocity is velocity of light in vacuum
--- $c$ \cite{HoG06}. The article is considered the information
system in 4-DPEIS, for 4-DPEIS is obtained the criterion of
invariant velocity existence. But before giving the formulation of
the theorem proving existence of information transference invariant
velocity is necessary made following rather important statement:

\begin{postulate}
\postulatecaption{existence of the minimal mean measured the
distance value in information space} The minimal mean measured the
distance value in information space is following value: $
\lambda_{c} \; = \; $ 1 bit \cite{She48,She49,She50,She51,She49MP,
Kol91}. \label{Post0}
\end{postulate}

The statement make \postulateref{Post0} was used at following
theorem proof:

\begin{theorem}
\theoremcaption{asserting necessary criteria of existence the
invariant velocity of the information transference}

There is the invariant velocity of the information transference -
{\it the light information velocity (LIV)} $ \nu_{c} $ (the unit of
the measure of the information velocity is bit per second) - which
is related to the minimal mean measured of the distance value in
information space $\lambda_{c}$ (the value of the minimal mean
distance in information space is established \postulateref{Post0})
and fundamental physical constants \cite{MoT05,Yao06}: the velocity
of light in vacuum -- $c$ , Planck's constant -- $\hbar$, the
constant of gravitational interaction -- $G_{N}$: \label{Ther1}
\begin{equation}
\begin{gathered}
\displaystyle{\nu_{c}} \; = \; \displaystyle{\sqrt{{{\lambda^{2}_{c}
\; c^{5}} \over {\hbar \; G_{N}}}}} \>, \mathtab
\displaystyle{\nu_{c}} \; = \; \displaystyle{1{.}6637 \times
2^{143}} \; \displaystyle{{bit}\over{c}}
\end{gathered}
\label{Cvel}
\end{equation}
\end{theorem}
\begin{proof} \proofcaption{Ther1}
For the theorem proof is considered the physical system in which the
electromagnetic wave extends with velocity of light in vacuum -- $c$
and the wave length to equal $\lambda \; = \; 2 \; \ell_P$, where
$\ell_P$ -- is the length of Planck\footnote{The consideration of
smaller length of a wave, for example, the value of the wave length
equal to value of Planck's length $\lambda\; = \; \ell_P$, does not
represent interest if considered logic of the information bit
perception. For perception of information bit necessary to apprehend
the physical signal (wave) equal to the semi-wave length, minimally
possible value of measured  which is equal to length of Planck in
physical system \cite{HoG06,Ple75,ScS77,AsC06}.}. That frequency of
the propagation wave is also invariant value $\nu_{p}$, this is
obvious:
\begin{equation}
\nu_{p} \; = \; {{c} \over {2 \; \ell_P \>}}, \label{nuP}
\end{equation}
Let by means of the electromagnetic wave is transferred some
information, presented in the form of a binary code, for simplicity
of reasoning and without restriction of the generality. In this
frame, what transferred the information containing in one bit
(according to statement made in postulate) is necessary transferred
and accepted one semi-wave, as for recognition of one the
information bit is enough accepted or transferred the semi-wave. In
the given example by virtue of smallness the chosen the wave length
conditionally is considered the point object --- the material point
is including in the quantity and the sense of the information. On
the basis of above considered is possible made the following
statement: physical transference of the information by means of the
electromagnetic wave answers displacement of the information some
bits number in information space; otherwise this statement is made
so: for moving one bit of the information is necessary and enough
moving the semi-wave in physical space. Hence the information
velocity is proportional to frequency half of \eqoref{nuP}:
\begin{equation}
\nu_{c} \; = \; {{c \; \lambda_{c}} \over {\ell_P}} \>,
\label{nuPtoC}
\end{equation}
If used formula for length Planck through fundamental physical
constants: velocity of light in vacuum -- $c$, Planck's constant --
$\hbar$ the constant of gravitational interaction -- $G$:
\begin{equation}
\displaystyle{\ell_P} \; = \; \displaystyle{\sqrt {{{\hbar \;G_{N}}
\over {c^{3}}}}} \>, \label{Lp}
\end{equation}
The having substitution the \eqoref{Lp} in \eqoref{nuPtoC}, gives
the following \eqoref{Cvel} for {\it LIV} value.

In the considered example all parameters of a transferred
electromagnetic wave were invariant, in any physical frame of
reference. By virtue of it is approved: the length of a semi-wave is
related with the information transferring in information space,
transfer with invariant velocity of \eqoref{Cvel}. The theorem is
proved.\label{Prof1}
\end{proof}

So, in conclusion of the paragraph, the general postulates are
formulated in 4-DPEIS for systems consideration and realization of
Poincare transformations group \cite{HoG06}:

\begin{postulate} \postulatecaption{substantiation of the 4-DPEIS introduction}
For any physical system (and not only physical) is entered related
with this system 4-DPEIS: three-dimensional information space in
which position of the information is set, and also time identically
equal to time in physical system, linked to a considered
intelligence system. The abstract 4-DPEIS entered in the article is
continuous, as on time coordinates, so on space coordinates.
Measured values are characterized by chance quantities, i.e. it is
impossible approving the given value authentically true value.
\label{Post1}
\end{postulate}
\begin{postulate} \postulatecaption{existence substantiations of inertial frame of references}
On the basis of \theoremref{Ther1} approve, that there is the
invariant velocity of the information transference in 4-DPEIS,
certain by the equality {\it LIV} --- $\nu_{c}$ in \eqoref{Cvel}. In
4-DPEIS are inertial frame of references, in which {\it LIV}
invariant value. \label{Post2}
\end{postulate}
\begin{postulate} \postulatecaption{substantiation of usage Poincare group transformations}
In 4-DPEIS is entered representation of Poincare group, where as
velocity of light in vacuum have used {\it LIV}. \label{Post3}
\end{postulate}

\section{Kinematics of the information in 4-DPEIS}~\label{SecVelosity}
There is no necessity detailed of consideration of the kinematics of
information description question in abstract 4-DPEIS. Let's note
only, that position of the information should be described by means
of covariant (or contravariant) the 4-vector\footnote{The
hereinafter the Greek letters, for example, the indexes is
$\alpha,\;\beta,$ etc., designate indexes with values 0, 1, 2, 3,
and Latin letters, for example, by indexes is $i, \; j,$ etc.,
designate indexes with values 1, 2, 3; besides for that after the
repeated dummy indexes occurring once superscript and once subscript
is supposed summation, except for the separate frames noted is meant
in the text , thus takes in the account pseudo-Euclidean metrics,
for the Greek letters from indexes.} of position -- $x^{\alpha}$ and
the corresponding 4-vector of velocity $v^{\alpha}$. The 4-vector
should be transformed with usage of Poincare group realization in
which as invariant velocity is used {\it LIV} -- $\nu_{c}$, at
transition from one frame of reference to other frame of
reference\cite{HoG06,Ple75,ScS77,AsC06}:
\begin{equation}
x^{\;\alpha} \; = \; \left( x^{0},\;
x^{1},\;x^{2},\;x^{3}\right)\;\>.~\label{chi}
\end{equation}
The relativistic invariance consists in scalar product preservation
at Poincare group transformations. At transition from one frame of
reference in other frame of reference of the position 4-vector
should be transformed on Poincare group representation, according
\postulateref{Post3} has following obvious form for a 4-vector
displacement transformation:
\begin{equation}
x^{\prime\;\alpha} \; =\; \Lambda^{\alpha}_{~\beta}\; x^{\;\beta}\;
+\;b^{\;\alpha}\>, ~\label{Pua}
\end{equation}
where $\Lambda^{\alpha}_{\beta}$ --- is Lorentz's transformation
matrix, $b^{\;\alpha}$ --- is  4-vector of time-space displacement
in 4-DPEIS.

In frame of if product of the 4-vector most on itself will turn out
the following equality for its component, following of requirements
of relativistic invariance is considered:
\begin{equation}
x^{2}\;=\;x^{\;\beta}\;x_{\;\beta}\; = \; \left(x^0\right)^{2}\;-\;
\left(x^1\right)^{2}\;-\; \left(x^2\right)^{2}\;-\;
\left(x^3\right)^{2} = \; \nu^{2}_{c} \; \tau^{2}\>,~\label{invX}
\end{equation}
where $x^{0}\;=\; \nu_{c}\; t$ --- is lights bit; $t$ --- is time in
the laboratory system of reference, in conformity
\postulateref{Post1} coincides in due course in physical system;
$\nu_{c} \; \tau$
--- is the 4-interval, invariant value; $\tau$
--- is intrinsic time which, according to \postulateref{Post1}
identically coincides with intrinsic time in considered physical
system. From the received obvious kind of a square of the 4-interval
in this frame have relation between infinitesimal time intervals in
intrinsic system of references and laboratory system of references
\cite{HoG06,Ple75,ScS77,AsC06}:
\begin{equation}
\nu^{2}_{c}\;\left(d\;\tau\right)^2\;= \;\nu^{2}_{c}\;
\left(d\;t\right)^2\; + \; \left(d\;x^{1}\right)^2\;+ \;
\left(d\;x^{2}\right)^2\;+ \;
\left(d\;x^{3}\right)^2\>.~\label{DintX}
\end{equation}
The considering, that the derivative of coordinate on time in a
laboratory system of reference is velocity of transference in the
laboratory system of reference:
\begin{equation}
{{d\;x^{i}}\over{d\;t}}\;=\; v^{i}\>,~\label{DXT}
\end{equation}
The definition using  of \eqoref{DXT} is received for infinitesimal
time intervals following relation \cite{HoG06}:
\begin{equation}
d\;\tau\;= d\;t\;\sqrt{1+ \; \left({{v^{1}}/{\nu_{c}}}\right)^2\;+
\; \left({{v^{2}}/{\nu_{c}}}\right)^2\;+ \;
\left({{v^{3}}/{\nu_{c}}}\right)^2} \>.~\label{DiX}
\end{equation}
The value of 4-velocity define as the derivative of the information
position 4-vector on intrinsic time \cite{HoG06}:
\begin{equation}
\begin{gathered}
v^{\alpha} \; = \;{{\displaystyle{d\;x^{\alpha}}} \over
{\displaystyle{d\;\tau}}}\; = \; {\displaystyle{\displaystyle{1}}
\over
\displaystyle{{\sqrt{1\;-\;\left(\displaystyle{v/{\nu_{c}}}\right)^{2}}}}}
\; \left(v^{0}, \; v^{1},\;v^{2},\;v^{3}\right)\>,\mathtab
v\;=\;\sqrt{\left(v^{1}\right)^{2}\; + \;\left(v^{2}\right)^{2}\; +
\;\left(v^{3}\right)^{2}}\>,~\label{vel}
\end{gathered}
\end{equation}
where $v^0\;\equiv\; \nu_{c}$ --- is the time component of the
4-velocity. The scalar product of the 4-velocity is received
following equation:
\begin{equation}
v^{\alpha}\;v_{\alpha}\; = \; {\displaystyle{\displaystyle{1}} \over
\displaystyle{{{1-\left(\displaystyle{v/{\nu_{c}}}\right)^{2}}}}}\;\left[
\left(v^0\right)^{2}-\; \left(v^1\right)^{2}\;-\;
\left(v^2\right)^{2}\;-\; \left(v^3\right)^{2}\right]\;
\equiv\;{\nu_{c}}^2\>,~\label{Pvel}
\end{equation}
It is possible to enter the following dimensionless 4-vector which
components are the  4-velocity attitude of \eqoref{vel} to value
$\nu_{c}$ obtain by \eqoref{Cvel}:
\begin{equation}
\begin{gathered}
\beta^{\;\alpha}\; = \;{\displaystyle{\displaystyle{v^{\,\alpha}}}
\over \displaystyle{\nu_{c}}}\; = \;{\displaystyle{\displaystyle{1}}
\over \displaystyle{{\sqrt{1\;-\;\beta^2}}}} \; \left( 1, \;
\beta^{1},\;\beta^{\,2},\;\beta^{\,3}\right)\>,\mathtab
\beta\;=\;\sqrt{\left(\beta^{\:1}\right)^{2}\; +
\;\left(\beta^{\:2}\right)^{2}\; +
\;\left(\beta^{\:3}\right)^{2}}\>,\mathtab
\displaystyle{\beta^{\;i}}=\displaystyle{{v^{\;i}}/{\nu_{c}}}\>.
\end{gathered}~\label{Bvel}
\end{equation}
The considering kinematics of the information is necessary to pay
attention to some contradiction arising by comparison of some
postulates, resulted in the paragraph of ~\ref{SecPostulate}. This
question is necessary to consider more in detail. The contradiction
is noticeably if to consider statements \postulateref{Post0} and
\postulateref{Post1}. The invariant distance existence on the one
hand affirms in information space, but on the other hand is
considered the continuous information space-time. What to resolve
this contradiction is necessary to pay attention to process of
measurement of physical values which is averaging of the measured
value \cite{She48,She49,She50,Kol91}. If to take for a basis is
received the given contradiction, the arising contradiction
permission  consists the mean values of the displacement 4-vector
space components would be proportional to integers setting
comparatively the invariant mean measured value \postulateref{Post0}
\cite{She48,She49,She50,Kol91}.

\begin{theorem}~\theoremcaption{of mean displacement in 4-DPEIS}
The invariant minimal displacement in information space, in any
inertial frame of reference is the value $\lambda_{0}\;=\;$ 1 bit
(it had according to the \postulateref{Post0}). Space components of
the 4-vector of mean displacement from the point $x_{a}^{\;\alpha}$
in the point $x_{b}^{\;\alpha}$ should is expressed through an
integer of bits,\mathtab ${\langle\Delta x^{\alpha}\rangle\;=\;
\left(\nu_{c}\;\langle\; \Delta\;
t\;\rangle,\;N_{1}\;\lambda_{0},\;N_{2}\;\lambda_{0},\;N_{3}\;\lambda_{0}\right)}$,
where $N_{1},\;N_{2},\;N_{3}$ --- are integers, à $\langle \Delta
t\rangle$
--- is the mean time interval for which there is the displacement.
By virtue of the given statement is received the following equality
for the 4-vector of displacement (these is received equation for
components of the given 4-vector) and 4-velocities: \label{Ther2}
\begin{equation}
\langle\Delta x^{\alpha}\rangle \; = \; \nu_{c}\;
\langle\displaystyle{\displaystyle{\int\limits_{x_{a}^{\;\alpha}}^{x_{b}^{\;\alpha}}}
{\beta^{\,\alpha}}} \;\;\; d\;\tau\rangle\>,~\label{V3}
\end{equation}
\begin{equation}
\begin{gathered}
N_{1} \; = \; \displaystyle{\nu_{c}\over\lambda_{0}}\;
\langle\displaystyle{\displaystyle{\int\limits_{x_{a}^{1}}^{x_{b}^{1}}}
{{\beta^{1}}\over\displaystyle{{\sqrt{1\;-\;\beta^{\;2}}}}}} \;\;\;
d\;\tau\rangle\>,\mathtab N_{2} \; = \;
\displaystyle{\nu_{c}\over\lambda_{0}}\;
\langle\displaystyle{\displaystyle{\int\limits_{x_{a}^{2}}^{x_{b}^{2}}}
{{\beta^{\;2}}\over\displaystyle{{\sqrt{1\;-\;\beta^{\;2}}}}}}
\;\;\; d\;\tau\rangle\>,\mathtab N_{3} \; = \;
\displaystyle{\nu_{c}\over\lambda_{0}}\;
\langle\displaystyle{\displaystyle{\int\limits_{x_{a}^{3}}^{x_{b}^{3}}}
{{\beta^{\;3}}\over\displaystyle{{\sqrt{1\;-\;\beta^{\;2}}}}}}
\;\;\; d\;\tau\rangle\>,\mathtab \langle\Delta t\rangle\; =
\displaystyle{\lambda_{0}^{2}\over \sqrt{\langle v^2\rangle}}\;
\sqrt{N^{\,2}_{1}\; + \; N^{\,2}_{2} \; + \;
N^{\,2}_{3}}\;\>,~\label{Vxyz}
\end{gathered}
\end{equation}
where the value $\langle v^2\rangle$ is an mean from $v^2$, certain
according to the \eqoref{vel}, and $\tau$ -- intrinsic time of
system.
\end{theorem}
\begin{proof}\proofcaption{Ther2} The proof of the given theorem has been lead to two steps.
The first have proved equalities for space components of the
displacement 4-vector. Without restriction of a generality have
considered displacement of the information along one axis, for
example is axis of $x^1 $. The velocity of displacement of the
information along this axis, at the relativistic description have
designated for convenience $u^1$:
\begin{equation}
u^1\;=\;\nu_{c}\;{{\beta^{\;1}}\over\displaystyle{{\sqrt{1\;-\;\beta^{\;2}}}}}\>.
~\label{TVel1}
\end{equation}
The displacement from a point $x_{a}^1$ in the point $x_{b}^1$
should be expressed by following integral from velocity along the
axis $x^1$:
\begin{equation}
\Delta\;
x^1\;=\;x^1_{b}-x^1_{a}\;=\;\int\limits_{x_{a}^{1}}^{\;x_{b}^{1}}u^1\;
d\;\tau\;=\;\nu_{c}\;\int\limits_{x_{a}^{1}}^{\;x_{b}^{1}}
{{\beta^{\;1}}\over\displaystyle{{\sqrt{1\;-\;\beta^{\;2}}}}}\;d\;\tau
\>.~\label{TBias1}
\end{equation}
The considering 4-velocities scalar product obvious form of
\eqoref{vel} and 4-velocity vector component of \eqoref{Bvel} at
statement of \eqoref{TBias1} is received following integral:
\begin{equation}
\Delta\; x^1\;=\;\int\limits_{x_{a}^{1}}^{\;x_{b}^{1}}v^1\;
d\;t\>,~\label{TBias2}
\end{equation}
where $t$-- is time in a laboratory frame of reference.

The continuous change of component of the displacement vector  along
the axis $x^1$ should be casual parameter (see
\postulateref{Post1}), therefore measured displacement should be
mean displacement calculated on rules of calculation of a continuous
random variable, not concretizing the form of the random variable
probability density (to the question of calculation of density
should be return later, considering dynamics of the information),
should be calculate an mean displacement:
\begin{equation}
\langle\Delta\;
x^1\rangle\;=\;\langle\;\int\limits_{x_{a}^{1}}^{\;x_{b}^{1}}v^1\;
d\;t\;\rangle\>.~\label{TAver1}
\end{equation}
By virtue of \postulateref{Post0} the measured invariant mean
displacement value is the value $\lambda_{0}\;=\;$ 1 bit, therefore
any mean displacement along any space axis in considered 4-DPEIS
should be proportional to the integer of bits. If to assume is
received contradiction, that $\lambda_{0}$ is not invariant value
that contradicts \postulateref{Post0}. Therefore is received, that
any mean displacement is proportional to the integers of bits:
\begin{equation}
\langle\Delta x^1\rangle\;=\;N_{1}\;\lambda_{0}\>.~\label{TAver2}
\end{equation}
From this equality in view of \eqsref{TBias1}{TAver2} should be
turned out demanded equality for component of $x^1$ \eqoref{Vxyz}:
\begin{equation} N_{1} \; = \;
\displaystyle{\nu_{c}\over\lambda_{0}}\;
\langle\displaystyle{\displaystyle{\int\limits_{x_{a}^{1}}^{x_{b}^{1}}}
{{\beta^{1}}\over\displaystyle{{\sqrt{1\;-\;\beta^{\;2}}}}}} \;\;\;
d\;\tau\rangle\>.~\label{Prov1}
\end{equation}
The analogously is possible to prove validity of the statement about
proportionality of components to the integer of bits for the
remained space components of the mean displacement vector in
\eqoref{Vxyz}.

The second stage of the proof should be the proof of a equality for
mean time, thus consider, that equalities for all space coordinates
are already proved by us. So, should be calculate mean value from
time component of a 4-vector of displacement:
\begin{equation}
\begin{gathered}
\Delta\; x^0\;=\;x_{b}^{0}-x_{a}^{0}\;\;=\;
\nu_{c}\;\int\limits_{x_{a}^{0}}^{\;x_{b}^{0}}{{1}\over\displaystyle{{\sqrt{1\;-\;\beta^{\,
2}}}}}\;d\;\tau\mathtab \Delta\;
x^0\;=\;\nu_{c}\;\left[t_{b}-t_{a}\right]\;=\;\nu_{c}\;\Delta t \>.
\end{gathered}~\label{TTime1}
\end{equation}
The mean time interval should be received as random value:
\begin{equation}
\langle\Delta x^0\rangle\;=\;\nu_{c}\;\langle\Delta t\rangle
\>.~\label{TTave1}
\end{equation}
The 4-interval definition of \eqoref{chi} took advantage, and the
4-interval should be calculated  for the mean displacement 4-vector
of \eqoref{V3}:
\begin{equation}
\left[\nu_{c}\;\langle\Delta
\tau\rangle\right]^2\;=\;\left[\nu_{c}\;\langle\Delta
t\rangle\;\right]^2\; - \;\left[ N_{1}\; \lambda_{0}\right]^2\;-
\;\left[ N_{2}\; \lambda_{0}\right]^2\;- \;\left[ N_{3}\;
\lambda_{0}\right]^2\>.~\label{Tinter1}
\end{equation}
This of \eqoref{Tinter1} is transformed and led to following form:
\begin{equation}
\left[\langle\Delta t\rangle^2\;-\;\langle\Delta
\tau\rangle^2\right]\;=\;\left({{\lambda_{0}}\over{\nu_{c}}}\right)^2\;\left[
N_{1}^2\;+ \;N_{2}^2\;+ \;N_{3}^2\right]\>.~\label{Tinter2}
\end{equation}
For calculation of mean values in the left part of \eqoref{Tinter2}
uses relate between time intervals in laboratory and intrinsic frame
of references (this frames of references is considering inertial
frame of references) \cite{HoG06,ScS77}:
\begin{equation}
\Delta \tau\;=\;\Delta t\;\sqrt{1\;-\;\beta^{\,2}}\>,~\label{Ttau}
\end{equation}
where $\beta$ the resultant equation for relate of the information
is certain by mean time of displacement \eqoref{Bvel}, as the result
of averaging the given equation and substitution of result in
\eqoref{Tinter2}:
\begin{equation}
\langle\Delta t\rangle\;=\;{{\lambda_{0}}\over{\sqrt{\langle
v^2\rangle}}}\;\sqrt{N_{\, 1}^2\;+ \;N_{\, 2}^2\;+ \;N_{\,
3}^2}\>.~\label{Tinter3}
\end{equation}
The \eqoref{Tinter3} coincides with equation for time component set
in \eqoref{Vxyz}. The theorem is proved. ~\label{Prof2}
\end{proof}

\section{Dynamics and parameters of the information}~\label{SecEmotion}
For the description of dynamics of the information should be
necessary made the assumption that key parameter of dynamics and
interaction is {\it the {\it GIE}} --- $Q$. The {\it GIE} parameter
is simultaneously analogue mass a physical body, i.e. acts as a
measure of inertia, and analogue of an electric charge, i.e. acts in
a role of a source of the field of interaction. The using this
assumption, for the description of the field generated by
interaction $Q$ on other information objects is possible to enter
formally 4-vector potential, having defined through a 4-interval and
4-velocity:
\begin{equation}
A^{\;\alpha} \; = \left(
A^{0},\;A^{1},\;A^{2},\;A^{3}\right)\>,~\label{Apot}
\end{equation}
where $A^{0}$--- is the time component, $A^{1},\;A^{2},\;A^{3}$
--- are space components of a 4-vector of potential which obvious
kind should be certain from conditions of a solved problem. The
notice is necessary , that components of the given vector depend on
parameter of the information transference and parameter of the
information interference, including from entered in the formal form
{\it the {\it GIE}}. What to calculate an obvious kind of components
of a 4-vector of interference potential was apply a technique of the
vector potential calculation used in electrodynamics, being based on
the made assumption. The considering this technique was enter the
differential operator whom was name 4-divergence, written out
accordingly in covariant and contravariant forms
\cite{HoG06,Ple75,ScS77,AsC06}:
\begin{equation}
\begin{gathered}
\partial_{\;\alpha} \; = \left(\displaystyle{{\partial}\over{\partial\;x^{0}}},
\;-\displaystyle{{\partial}\over{\partial\;x^{1}}},\;-\displaystyle{{\partial}\over{\partial\;x^{2}}},
\;-\displaystyle{{\partial}\over{\partial\;x^{3}}}\right)\>,\mathtab
\partial^{\;\alpha} \; = \left(\displaystyle{{\partial}\over{\partial\;x_{0}}},
\;\displaystyle{{\partial}\over{\partial\;x_{1}}},\;\displaystyle{{\partial}\over{\partial\;x_{2}}},
\;\displaystyle{{\partial}\over{\partial\;x_{3}}}\right)\>.
\end{gathered}~\label{diver}
\end{equation}
The vector potential $A^{\alpha}$ should be used allowed to
construct antisymmetric the second rank tensor
\cite{HoG06,Ple75,ScS77,AsC06}:
\begin{equation}
G^{\;\alpha\;\beta} \; = \; \left(
\partial^{\;\alpha}\;A^{\;\beta}\; -\;
\partial^{\;\beta}\;A^{\;\alpha}\right)\>,~\label{Gdu}
\end{equation}
The antisymmetric tensor of \eqoref{Gdu} has allowed to construct
dual tensor, by a following rule:
\begin{equation}
\displaystyle{\tilde{G}^{\;\alpha\;\beta} \; =
\;\displaystyle{{1\over 2}\;}
\displaystyle{\varepsilon^{\;\alpha\;\beta\;\varrho\;\sigma}}\;
\displaystyle{G_{\;\varrho\;\sigma}}}\>,~ ~\label{Gddu}
\end{equation}
where $\varepsilon$ --- antisymmetric 4 ranks Levy-Chivita tensor
\cite{HoG06,Ple75,ScS77,AsC06}. By analogy to the description of the
electromagnetic field is possible formally has been written out the
set of equations describing generated by the {\it GIE} which in this
frame acted as the electric charge. The current density was enter
formally 4-dimensional vector \cite{HoG06,Ple75,ScS77,AsC06}:
\begin{equation}
j^{\;\alpha}\;= \; \left(\;j^{\;0}_{\rho},
\;j^{\;1}_{\rho},\;j^{\;2}_{\rho},\;j^{\;3}_{\rho}\right)\>,~\label{Qtok}
\end{equation}
Which components are in detail considered further. That considered
$Q$ by virtue as the charge could formally have demanded from
performance of the the charge preservation law, in this frame
received for a 4-dimensional vector of current density following
formal equation \cite{HoG06,Ple75,ScS77,AsC06}:
\begin{equation}
\partial^{\;\alpha}j_{\;\alpha}\;=\;0\>.~\label{sav}
\end{equation}
The formal set of equations is as a result written out, relating
vector 4-potential with the source of the field generated by this
potential, presented by formally entered 4-vector of current density
\cite{HoG06,Ple75,ScS77,AsC06}. By analogy to electrodynamics have
received set of Maxwell's equations, including, that as well as in
electrodynamics of component of a 4-vector of potential consider as
the contributions caused by presence of sources (i.e. is the {\it
GIE}), and the contributions which are not having such sources, and
describing movement of the specified sources in information space:
\begin{equation}
\left\{
\begin{gathered}
\displaystyle{ 1\over
2}\;{\partial_{\;\alpha}\;G^{\;\alpha\;\beta}\;=\;4
\pi\;j^{\;\beta}}\>,\mathtab \displaystyle{ 1\over
2}\;\partial_{\;\alpha}\;\tilde{G}^{\;\alpha\;\beta}\;=\;\;\;\;0~~~\>.
~\label{Mac}
\end{gathered}\right.
\end{equation}
The first equation in the given system has allowed to specify in an
obvious kind of component of a 4-vector of the potential, linked
with sources of the generated field a source $Q$, the second
equation has allowed to describe components of a 4-vector of
potential for which there is no source. The such formal introduction
and written out  equations Maxwell by analogy to electrodynamics
system made properties of a 4-vector of potential and a 4-vector of
current density are in detail considered.

On the 4-vector of potential are adjusted following requirements:
first, this requirement of invariancy of potential of interference
for any frame of reference; in the second, this requirement of
decrease of potential eventually and depending on remoteness from a
source in space, in a basis of this requirement is necessary
supervision that interaction of the information weakens depending on
increase in a time interval and its remoteness from a source of the
information. On the basis of these requirements the following
statement, potential inversely proportional to a 4-interval certain
by \eqoref{invX}, is made, and is directly proportional  $Q$ as to a
source of the field, in this frame the potential is presented by
following equation:
\begin{equation}
\begin{gathered}
\displaystyle{A^{\alpha} \; = \;{{Q}\over{x^{2}}}\;
a^{\;\alpha}}\>,\mathtab \displaystyle{A^{\alpha} \; =
\;{{Q}\over{x^{2}}}\; \left(
a^{0},\;a^{1},\;a^{2},\;a^{3}\right)}\>,
\end{gathered} ~\label{Abpot}
\end{equation}
where $a^{0},\;a^{1},\;a^{2},\;a^{3}$ --- are dimensionless
components which obvious forms defines from the decision of the set
of equations \eqoref{Mac}.

The having considered, the operator a 4-vector of current density of
\eqoref{Qtok}, have received, that as the {\it GIE} --- $Q$ borrows
4-volume by virtue consider density of the {\it GIE} follows as
4-density:
\begin{equation}
 \rho_{Q} =
{{d\;Q}\over{d^{\;4}\;x}}\>,~\label{Plot}
\end{equation}
where $d^{\;4}\;x$ --- is a differential 4-volume of information
space:
\begin{equation}
 d^{\;4}\;x\; =
\;d\;x^{0}\;\;d\;x^{1}\;\;d\;x^{2}\;\;d\;x^{3}\>,~\label{Vol4}
\end{equation}
On the basis of definition of 4-density of the {\it GIE} is
redefined the 4-vector of current density, having selected obviously
dependence on 4-density of the {\it GIE}:
\begin{equation}
j^{\;\alpha}\;= \; {{d\;Q}\over{d^{\;4}\;x}}\left(\;j^{\;0},
\;j^{\;1},\;j^{\;2},\;j^{\;3}\right)\>,~\label{tok}
\end{equation}
where $j^{\;0}, \;j^{\;1},\;j^{\;2},\;j^{\;3}$ --- is dimensionless
components of a 4-vector of current density which obvious forms
should be certain from the decision of the equations set
\eqoref{Mac}.

The now have considered information dynamic parameter have entered
the analogue of 4-vector of momentum for the information system. For
this purpose  4-velocity definition of \eqoref{vel} and the value of
the {\it GIE} entered earlier by us is used:
\begin{equation}
p^{\;\alpha} \; =\; Q\; v^{\;\alpha}\>.~\label{Mom}
\end{equation}
Through introduced by us value potential 4-vector of \eqoref{Abpot}
and the 4-vector of momentum  \eqoref{Mom}, have written out
Lagrangian systems. First appointed Lagrangian for free information
system in form \cite{HoG06,ScS77}:
\begin{equation}
\mathfrak{L}_{\;0}\left(x^{\alpha},v^{\alpha}\right) = -
p^{\;\alpha}\; v^{\alpha} =-\;Q\; v_{\alpha}\; v^{\alpha}=
-Q\;\nu^{\;2}_{c}\>.~\label{LagF}
\end{equation}
The sort of Lagrangian is dictated by a requiring of minimality
Lagrange action integral determine as integral on intrinsic time
from Lagrangian at the displacement of the information from the
point $x_{a}^{\alpha}$ in the point $x_{b}^{\alpha}$ information
space \cite{HoG06,ScS77}:

\begin{equation}
S_{0}\left(x_{a}^{\alpha},x_{b}^{\alpha}\right) =\int
\mathfrak{L}_{\;0}\left(x^{\;\alpha},v^{\;\alpha}\right)\;d\;\tau =
-\int Q\; v_{\alpha} \;d\;x^{\alpha}\>.~\label{SFdef}
\end{equation}

If movement of the information, described by the {\it GIE} ---
$Q_{1}$ was implemented under interaction of other information,
described the {\it GIE} --- $Q_{2}$, then Lagrangian has accepted
resulting form:
\begin{equation}
\mathfrak{L}_{\;int}\left(x^{\;\alpha},v^{\;\alpha}\right) =
-\left(Q_{1}+
Q_{2}\right)\left[\;f\;v_{\;\alpha}+A_{\;\alpha}\;\right]v^{\;\alpha}\>,~\label{LagP}
\end{equation}
where the dimensionless multiplier $f$ was introduced for
convenience of a designation and equal:
\begin{equation}
f = {{1}\over{2}}\left[ 1 + {{Q_{1} - Q_{2}}\over{Q_{1} +
Q_{2}}}\;{{u^{\;\alpha}\;v_{\;\alpha}}\over{\nu^{2}_{c}}}\right]\>,~\label{Ffacdif}
\end{equation}
and the 4-velocities $v^{\;\alpha}$ and $u^{\;\alpha}$ were defined
in velocities of the first information $v^{\;\alpha}_{1}$ and the
second information $v^{\;\alpha}_{2}$ in a following form:
\begin{equation}
\left\{
\begin{gathered}
{v^{\;\alpha} = \left(v^{\;\alpha}_{1}+
v^{\;\alpha}_{2}\right)}\>,\mathtab {u^{\;\alpha} =
\left(v^{\;\alpha}_{1}- v^{\;\alpha}_{2}\right)}\>,\mathtab
\end{gathered}
\right.
~\label{Vel12}
\end{equation}
The 4-vector of potential $A^{\;\alpha}$ interferences in this frame
was defined by following equation:
\begin{equation}
A^{\;\alpha} = {{Q_{1 2}} \over {x^{\;2}}}\; \left(
a^{0},\;a^{1},\;a^{2},\;a^{3}\right)\>,~\label{Aint}
\end{equation}
where the {\it GIE} used potentially $Q_{1 2}$ was expressed through
each information emotions  $Q_{1}$ and $Q_{2}$:
\begin{equation}
Q_{1 2} = {{Q_{1}\;Q_{2}}\over{\left(Q_{1}+
Q_{2}\right)}}\>,~\label{Qint}
\end{equation}
Using equations for Lagrangian interferences of information $Q_{1}$
and $Q_{2}$ of \eqoref{LagP}, in view of \eqsref{Ffacdif}{Qint} for
Lagrange action integral have got resulting obvious form:
\begin{equation}
\begin{gathered}
S_{\;int}\left(x_{a}^{\alpha},x_{b}^{\alpha}\right)
=\displaystyle{\int
\mathfrak{L}_{\;int}\left(x^{\;\alpha},v^{\;\alpha}\right)\;d\;\tau}
\>,\mathtab S_{\;int}\left(x_{a}^{\alpha},x_{b}^{\alpha}\right) = -
\displaystyle{\int \left(Q_{1}+
Q_{2}\right)\left[\;f\;v_{\;\alpha}+A_{\;\alpha}\;\right]
\;d\;x^{\alpha}}\>.
\end{gathered}
~\label{Sdef}
\end{equation}
The having written out, thus Lagrangian interferences of
information, the question on dimension of the {\it GIE} and a
component of a 4-vector of potential of interference is solved. Was
plainly, that components of a 4-vector of potential of interference
should index on dimension with the information 4-velocity
of\eqoref{LagP} -- [bit/c]. The obvious form of the interference
potential 4-vector of \eqoref{Abpot} indicates on that fact, that
dimension a component of a 4-vector of potential of interference
inversely proportional the bit in a square -- [bit$^2$]. Hence,
dimension of the {\it GIE} is got -- [bit$^3$/c], i.e. the
three-dimensional volume of the information in unit of time, thus
indicates a meaning of the {\it GIE} on that the given information
is perceived, or not. The problem having considered  on dimension of
the {\it GIE}, it is possible to consider the problem on existence
of the invariant {\it GIE} and about its relation with any other
{\it GIE}. The {\it GIE} admissible meanings question treating was
do on the basis of that the given value is the characteristic of
interaction. The {\it GIE} is similar to a charge, it can differ a
sign (thus it is supposed, that the same emotions are drawn, and
heteronymic make a start, has been more in detail view this question
below). The measure of the {\it GIE} invariant existence  and mutual
relation with any other meaning of emotion is the following theorem:
\begin{theorem}
\theoremcaption{asserting {\it GIE} -- $Q_{c}$ dimension constant
existence and restriction criterions} There is constant measured
value in 4-DPEIS the minimal mean distance $\lambda_{c}$, hence,
there is a constant measured minimal mean volume $\lambda^{3}_{c}$,
means it is possible to define a constant, on dimension indexing
with the introduced {\it GIE}, and to express it through
$\lambda_{c}$ fundamental manual constants \cite{MoT05,Yao06}:
velocity of light in vacuum -- $c$, Planck's constant -- $\hbar$,
the constant of gravitational interaction -- $G_{N}$. The {\it GIE}
dimension constant have obvious form: \label{Ther3}
\begin{equation}
\begin{gathered}
\displaystyle{Q_{c}}\; = \displaystyle{\sqrt{{{\lambda^{6}_{ñ}\;
c^{5}}\over{\hbar\;G_{N}}}}}\>,\mathtab \displaystyle{Q_{c}\; =
\;1{.}6637\times
{2^{143}}}\;\displaystyle{{\hbox{bit}^{3}}\over{c}}\>.
\end{gathered}~\label{Qcdef}
\end{equation}
Thus any {\it GIE} where following designations are entered for each
information parameters with the $Q_{1}$ and $Q_{2}$ -- are the {\it
GIE} and also them combinations of these parameters corresponding:
gathered in invariant unit volume $\lambda^{3}_{c}\;=$ 1 bit$^3$,
apprehended for some time interval, it is related with a constant of
dimension of the {\it GIE} a following inequality:
\begin{equation}
\left|Q\right| \leqslant Q_{c}\>.~\label{NerQcdef}
\end{equation}
\end{theorem}
\begin{proof}\proofcaption{Ther3} The proof of the theorem are conducted in two stages the question
on existence of an invariant value, on the dimension indexing with
emotion, first, is viewed. As there is an mean minimal distance
$\lambda_{c}$, accordingly if three space measurements along each of
coordinates axes received have been taken, that the considered
volume $\lambda{3}_{c}$ will be too constant. If the given volume
has been apprehended for a time interval equal Planck's to time
$t_{P}$, the following value certain as result:
\begin{equation}
\displaystyle{Q_{c}\; = \; {{\lambda_{c}^{3}}\over
{t_{P}}}}\>,~\label{QLdef}
\end{equation}
Was plainly, should be a constant in any system of readout, on
dimension will index with introduced before the {\it GIE} (see, for
example, a determination Lagrangian of free transference information
of \eqoref{LagF} and Lagrangian of interference information of
\eqoref{LagF} in information space).

The considering fundamental physical constants: velocity of light in
vacuum $c$, Planck's constant $\hbar$, gravitational constant $G$
can be made equation for Planck's time:
\begin{equation}
t_{P} \; = \;
\displaystyle\sqrt{{{\hbar\;G_{N}}\over{c^5}}}\>,~\label{Tp}
\end{equation}
The substituting \eqoref{Tp} in \eqoref{QLdef} have received the
demanded equation for {\it GIE} of \eqoref{Qcdef}.

Secondly, the {\it GIE} concluded in invariant unit 3-volume, equal
$ \lambda^{3}_{c}$ of \eqoref{QLdef} has been considered, let could
be apprehended for some any time interval $ \Delta \; t $, in this
frame emotion could be equal, including on the module:
\begin{equation}
\displaystyle{\left|Q\right|\; = \; {{\lambda_{c}^{3}}\over
{\Delta\;t}}}\>.~\label{QrecI}
\end{equation}
If the following limit could be calculated:
\begin{equation}
\displaystyle{\lim\limits_{\Delta\;t\rightarrow t_{P}}
\left|Q\right|\; = \; \lim\limits_{\Delta\;t\rightarrow
t_{P}}{{\lambda_{c}^{3}}\over {\Delta\;t}}}\; =
\;Q_{c}\>.~\label{QrecII}
\end{equation}
The apparently in limit the value of size received at the first
stage of the proof has turned out. The {\it GIE} inversely
proportional on a time interval received, that with increase in a
time interval value of {\it GIE} decreases. The Planck's time
$t_{P}$ --- is minimal measurable physical time interval, there
$Q_{c}$ --- is the greatest possible value {\it GIE} concluded in
unit 3-volume $\lambda^{3}_{c}$. Thus, received validity of the
statement \eqoref{NerQcdef}. The theorem is proved.~\label{Prof3}
\end{proof}

The {\it GIE} dimension constant existence criterion established in
\theoremref{Ther3}, and also relate with the {\it GIE} concluded in
unit 3-volume $\lambda^{3}_{c}$, has allowed to make following
transformations to considered potential. The potential was sent
dimensionless values. The {\it GIE} definition at a 4-vector of
potential of \eqoref{Abpot} have defined as follows, through
constant of {\it GIE} of \eqoref{Abpot}:
\begin{equation}
Q\; = \;Q_{c}\; n\left( x^{\;\alpha}\right)\;q\left(
x^{\;\alpha}\right)\>,~\label{Qqdef}
\end{equation}
where $n\left(x^{\;\alpha}\right)$ ---  is the number of elementary
volumes $\lambda_{c}^{3}$ which are borrowed with the given
information in information space; $q \left(x^{\;\alpha}\right)$ -
there is the dimensionless function describing perception of the
information on value. This function only one requirement, it should
be invariant of concerning Poincare group transformations. The
square of a 4-interval was entered following definition:
\begin{equation}
x^2\; = \;\lambda^{2}_{c}\; \left[s\left(
x^{\;\alpha}\right)\right]^{\,2}\>,~\label{hxdef}
\end{equation}
where $s \left( x^{\; \alpha} \right) $ - the dimensionless function
numerically equal to value of a 4-interval. On the basis of the
given transformations of \eqoref{Qqdef} and \eqoref{hxdef} is
received, the 4-vector of potential gets demanded dimension in
obvious form:
\begin{equation}
\displaystyle{A^{\alpha} \; = \;\nu_{c}\;n\left(
x^{\;\alpha}\right)\;{{q\left( x^{\;\alpha}\right)} \over
{\left[s\left( x^{\;\alpha}\right)\right]^{\,2}}}\;
a^{\;\alpha}}\>,~\label{AQpot}
\end{equation}
The 4-vector of momentum has similarly redefined, using
\eqoref{Bvel} and \eqoref{Qqdef}:
\begin{equation}
p^{\,\alpha} \; = \;Q_{c}\;\nu_{c}\;n\left(
x^{\;\alpha}\right)\;q\;\left(
x^{\;\alpha}\right)\;\beta^{\,\alpha}\>,~\label{Pbpot}
\end{equation}
The result, for Lagrangian of free transference information of
\eqoref{LagF} and Lagrangian of interference information of
\eqoref{LagP} were received following equations:
\begin{equation}
\begin{gathered}
\mathfrak{L}_{\;0} = - \;Q_{c}\;\nu^{2}_{c}\;\left[n\left(
x^{\;\alpha}\right) \;q\left(
x^{\;\alpha}\right)\;\beta^{\,\alpha}\;\beta_{\,\alpha}\right]\>,\mathtab
\displaystyle{\mathfrak{L}_{\;int} = -\;Q_{c}\;\nu^{2}_{c}\;\left(
n_{+}\;q_{+} \left[f\;\beta_{\;\alpha\;+}+{{q_{1 2}} \over
{\left[s\left( x^{\;\alpha}\right)\right]^{\,2}}}\;
a^{\;\alpha}\right]\beta^{\;\alpha}_{\;+}\right)}\>,
\end{gathered}~\label{LagB}
\end{equation}
where following designations are entered for each information
parameters with the $Q_{1}$ and $Q_{2}$ --- are the {\it GIE} and
also them combinations of these parameters corresponding:
\begin{equation}
\displaystyle{\begin{gathered} n_{+}\;=\;n_{1}\left(
x^{\;\alpha}\right)\;n_{2}\left( x^{\;\alpha}\right)\>,\mathtab
q_{+} \;=\; \displaystyle{{1}\over{n_{2}}}\;q_{1}\left(
x^{\;\alpha}\right)\; +\;\displaystyle{{1}\over{n_{1}}}\;q_{2}\left(
x^{\;\alpha}\right)\>,\mathtab q_{-}\;
=\;\displaystyle{{1}\over{n_{2}}}\; q_{1}\left(
x^{\;\alpha}\right)\; -
\;\displaystyle{{1}\over{n_{1}}}\;q_{2}\left(
x^{\;\alpha}\right)\>,\mathtab q_{1
2}\;=\;\displaystyle{{q_{1}\left( x^{\;\alpha}\right)\;q_{2}\left(
x^{\;\alpha}\right)}\over{q_{+}\left(
x^{\;\alpha}\right)}}\>,\mathtab f \;=\;
\displaystyle{{{1}\over{2}}\;\left[ 1 \;+\;
q_{\pm}\;{\beta_{-}^{\;\alpha}\;\beta_{\;\alpha\;+}}\right]}\>,\mathtab
q_{\pm}\;=\;\displaystyle{{q_{-}}\over{q_{+}}}\>,\mathtab
\beta^{\;\alpha}_{+}\;=\; \beta^{\;\alpha}_{1}\; +\;
\beta^{\;\alpha}_{2}\>,\mathtab \beta^{\;\alpha}_{-}\;=\;
\beta^{\;\alpha}_{1}\; -\; \beta^{\;\alpha}_{2}\>.
\end{gathered}}~\label{LagBD}
\end{equation}

\begin{theorem}\theoremcaption{asserting existence of the information constant $\hbar_{c}$}
The information constant $\hbar _ {c}$ value is expressed through
fundamental physical constants \cite{MoT05,Yao06}, such as Planck's
constant - $\hbar$, velocity of light in vacuum - $c$, with a
constant of gravitational interaction - $G_{N}$ and constant of the
minimal average displacement in information space $\lambda _ {c}$ in
the following form: \label{Ther4}
\begin{equation}
\begin{gathered}
\hbar_{c} \; = \; \displaystyle{{\lambda_{c}^{5}\;
c^{5}}\over{2\;\pi\;\hbar\; G_{N}}}\>,\mathtab \hbar_{c} \; =
\;\displaystyle{1{.}7621\times {2^{284}}}\;
\displaystyle{{\hbox{bit}^{5}}\over{c^{2}}}\>.
\end{gathered}
~\label{hdef}
\end{equation}
\end{theorem}
\begin{proof}\proofcaption{Ther4}
For the proof of the given theorem has been considered the analogy
to physical system. As is known from quantum physics, if in this
frame was considered a particle in intrinsic frame of reference, for
rest elementary particle with nonzero rest mass was possible
according to representation about wave-corpuscular dualism
\cite{Aul05}, to write down the following equation:
\begin{equation}
\hbar\; \omega \; = \; \displaystyle{m\; c^{2}}\>,~\label{hom}
\end{equation}
where $\omega$ -- is a cyclic frequency, $m$ -- is rest mass of a
particle. Thus, have received, that on dimension Planck's constant
$\hbar$ is directly proportional to mass of a particle, velocity of
light in vacuum and time, since inversely proportional to cyclic
frequency. This analogy is used at calculation of information
constant. That was received by virtue the information constant
should be expressed through a constant of emotion, the {\it LIV} and
Planck's time:
\begin{equation}
\hbar_{c}\; = \;
\displaystyle{{1}\over{2\;\pi}}\;\displaystyle{Q_{c}\;
\nu_{c}^{2}}\;t_{P}\>.~\label{hcom}
\end{equation}

The fundamental physical constants have considered velocity of light
in vacuum $c$, constant Planck $\hbar$, gravitational constant
$G_{N}$ of which the obvious form for Planck's time of \eqoref{Tp}.
The substituting \eqoref{Tp} for Planck's time has been made an
obvious forms by the constant {\it GIE} of \eqoref{Qcdef} and
\eqoref{Cvel} for {\it LIV} in the received of \eqoref{hcom}. The
relate between information constant and fundamental physical
constants have received of \eqoref{hdef}. The theorem is
proved.~\label{Prof4}
\end{proof}

It will be in summary rebuked about that the considered way of the
account of interference not unique. The interaction inclusion will
be possible to use offered P.A.M. Dirac ways of direct interaction
inclusion in algebra observable. The P.A.M. Dirac article
\cite{Dir49} has offered three most economical ways of direct
interaction inclusion in the observable algebra which have received
the name of dynamics forms: light front, instant and point. Forms of
dynamics differ observable including interaction and kinematics
subgroups (observable independent from interaction). The application
of such approach on inclusion of interaction can be useful by
consideration of the general properties of systems, in particular at
construction of theories of dynamics and interference of the
information, under conditions not preservations of a charge or the
description of the related conditions
\cite{Pol89,KeP91,Lev93,Lev95}. In particular, the technique
developed within the limits of the instant form of dynamics
\cite{KrT93,BaK96,Kru97,BaK00,KrS01} can be adapted for the
description of information interactions.

\section{The GIE structure}
The one of key model elements is formally entered {\it GIE}
\cite{Tat89,Koz91,LoB05,FaB06,Bur06,Per07,Zie07,Mik07}, which
simultaneously represents the measure of the information inertia
({\it GIE} acts in a mass role if to result physical analogy, at
definition of kinematics values) and the measure of interaction
({\it GIE} acts in a charge role at definition of interaction
potential 4-vector analogue). Such formal introduction later is
validated, for this purpose are discussed contributions to {\it GIE}
of the information various components:{\it the text} --- is $\mu$,
the contribution given {\it text} in {\it context} --- is $\psi$ and
{\it implied sense} --- is $\gamma$, resulting interaction of the
given text with a context
\cite{Tat89,Koz91,LoB05,FaB06,Bur06,Per07,Zie07,Mik07}. The later on
the {\it GIE} dimensionless part --- $q$, accordingly is considered
and all contributions to the {\it GIE} were considered as
dimensionless value:
\begin{equation}
q\;=\;\mu\;+\;\gamma\;+\;\psi\>,~\label{qmgpdef}
\end{equation}
The basis of definition, for the generalized information emotion
\eqref {Qqdef} through a dimensionless part $q $ which is presented
in the paragraph \ref{SecEmotion}, the statement is possible make,
according to output \theoremref{Ther3}, that is defined by following
restriction $\left|q\right|\leq 1 $. The notice is necessary , that
in this frame it is possible to attribute conditional values $q$ in
which have been reflected logic character of values $q$: $q\;=\;1$ -
value {\it true} is attributed, $q\;=\;-1 $ - value is attributed
{\it false}, $q\;=\;0$ - is attributed value {\it uncertain}, i.e.
not {\it true} and not {\it false}.

For what to enter function of emotion, it was necessary consider the
information representation. There are two basic ways of the
information representation: or in the form of entropy source
\cite{She48,She49,She50,She51,She49MP,Kol91,Haz98,AsZ07,JoV07,HiM07,Mas07};
or in the logic functions form and the operations certain above them
\cite{CoW05,Ben05,Kur06,Zad65,Zad68,LeZ69,Zad75I,Zad75II,Zad05,GaP04,Qiu05,PeR06,SeC07,HuC07,DeC07}.
The complexity of such definitions usage consists: first, this was
difficult considered subjects properties of the information
generation or perception (the subjects properties was depended
correctness information generation or perception), secondly, in
approach of Shannon \cite{She48,She49,She50} was difficultly
considered the information logic component, thirdly, in
structurally-logic approaches, in particular, in fuzzy logics
\cite{Zad65,Zad68,LeZ69,Zad75I,Zad75II,Zad05,GaP04,Qiu05,PeR06,SeC07,HuC07,DeC07},
was complicated consideration of questions on generation and the
information dynamics.

In the given article is offered set the information in the any rank
tensor form, which components are responsible for the information
representation in the initial characters form (in simple frame is
the letter), the text symbolical designs (symbolical designs is the
words having certain value in the given text), the basic ways of
construction of the information structure  (grammatical and semantic
rules on which are under construction the basic offers), rules of
construction and perception of separate phrases, etc. notice at once
is necessary, that given way of the information job is not
unequivocal, but allows to consider not only features of the given
the information representation, but also subjective features as
information and perception of this information.

By consideration of the text contribution is necessary for us to
consider three objective circumstances. First, the text generally
will present in any rank tensor form \cite{She49MP,ChR07,LeB07}:
components of data tensor contain signs on the text (for example
letters), features of its punctuation, author's style, etc. If
without the generality restriction was considering not the verbal
level then tensor components contain certain characters perceived
entirely (for example inarticulate sounds or emotional gestures). At
transition from one frame of reference in other frame of reference
the tensor of text will be transformed under the following law, that
is had 4$^{m}$ component:
\begin{equation}
\tensor{T}{\alpha_{1}}{\alpha_{2}}{\alpha_{m-1}}{\alpha_{m}}
{\beta_{1}}{\beta_{2}}{\beta_{m-1}}{\beta_{m}}\>.~\label{text}
\end{equation}
Secondly, for the text perception is necessary to possess the tool
on perception \cite{BoR96,HeB01BCFGM,RiB01,HeB01,BoR05,MoC07}. As
such tool can act tensor the text perception which components
comprise signs which probably to apprehend, rules of perception of
the text, etc. (i.e. characterizes features of the subject of the
information perception
\cite{BoR96,HeB01BCFGM,RiB01,HeB01,BoR05,MoC07}). The dimension of
tensor of text perception should be not less than the perceived
text, i.e. 4$^{m}$-component. At transition from one frame of
reference in other frame of reference of transformation occur by
rules of tensor transformation:
\begin{equation}
\tensor{D}{\alpha_{1}}{\alpha_{2}}{\alpha_{m-1}}{\alpha_{m}}
{\beta_{1}}{\beta_{2}}{\beta_{m-1}}{\beta_{m}}\>.~\label{imput}
\end{equation}
Thirdly, interaction between two various texts was carried out as
interaction between scalars in the chosen point in 4-DPEIS, on the
basis can defined scalar value {\it the text emotion}, such value
can be carried out by convolution initial tensor of text with tensor
of text perception and averaging on number of tensor of text
components:
\begin{equation}
\mu=\tensum{D}{T}{\alpha_{1}}{\alpha_{m}}\>,~\label{mudef}
\end{equation}
where $\mu$ --- is {\it the text emotion}; $Q_{c}$ - is the
generalized information emotion dimension constant, which value of
\eqoref{NerQcdef} is certain according to \theoremref{Ther3}; $n
\left(x^{\;\alpha}\right)$ - is elementary volumes number borrowed
by the information in information space. The two tensors general
property should be such that at transition from one frame of
reference in 4-DPEIS in other system {\it the text emotion} remained
invariant value for the given concrete point in 4-DPEIS.

The text and text perception is characterized not only tensors of
\eqoref{text} and \eqoref{imput}, but also change and heterogeneity
component in 4-DPEIS. For that what to consider the text changes and
heterogeneity, could be entered following text deformations tensor
which is defined by the operator - tensor heterogeneity of the text
on components with an index $\nu$:
\begin{equation}
\Tdiv {\nu}{T}{\alpha_{1}}{\alpha_{m}}{T}\>.~\label{Tdivdef}
\end{equation}
The dimension received tensor of \eqoref{Tdivdef} differs from
tensor of text \eqoref{text}, tensor of \eqoref{Tdivdef} contains
4$^{n-1}$ component. These perceived changes is possible only having
processed changes of the text by the device means the text
perception as which is used the tensor of text perception, as
tensors of \eqoref{imput} and \eqoref{Tdivdef} convolution result is
received some 4-vector value which influences the given text
perception:
\begin{equation}
\begin{gathered}
\streamF{\nu}{T}{D}{\alpha_{1}}{\alpha_{m}}{I} \mathtab
\streamS{\nu}{T}{D}{\alpha_{1}}{\alpha_{m}} \>.
\end{gathered}~\label{Idef}
\end{equation}
The index $\nu$ in \eqoref{Idef} runs all possible values from
$\alpha_{1}$ up to $\alpha_{n}$ inclusively. The \eqoref{Idef} is
averaging on all possible changes and heterogeneity component in
4-DPEIS. The besides changes and heterogeneity text might be
inherited changes and heterogeneity components of tensor of text
perception, acting on analogy with tensor of texts, could enter
tensor of the text heterogeneity perception on index $ \nu $:
\begin{equation}
\Tdiv {\nu}{D}{\alpha_{1}}{\alpha_{m}}{D}\>.~\label{Ddivdef}
\end{equation}
The entered tensor of \eqoref{Ddivdef} by means, at convolution with
tensor of \eqoref{text} and the subsequent averaging on all possible
values of components with every possible values of an index $\nu$,
the perceived text distortions 4-vector, caused by the text
perception changes related with changes and heterogeneity tensor of
text perception has turned out as a result. It is supposed, that all
changes and heterogeneity of the tensor of text perception are
related to influence on context, the received vector - will refer to
as a vector of context stream:
\begin{equation}
\begin{gathered}
\streamF{\nu}{D}{T}{\alpha_{1}}{\alpha_{m}}{B}\mathtab
\streamS{\nu}{D}{T}{\alpha_{1}}{\alpha_{m}}\>.
\end{gathered}~\label{Bdef}
\end{equation}

The tensors of \eqoref{Tdivdef} and \eqoref{Ddivdef} also by
facility is defined one more scalar value which is referred to as
the basic the text to context contribution:
\begin{equation}
\begin{gathered}
\PsidevF{\psi}{{0}}{T}{D}{\alpha_{1}}{\alpha_{m}}\mathtab 
\PsidevS{D}{T}{\alpha_{1}}{\alpha_{m}}\>.
\end{gathered}~\label{Psdef}
\end{equation}

The implied sense on the one hand and the text at interaction with
context cannot exist the friend without the friend, in other words
the text at interaction with context generates implied sense or on
the contrary appeared implied sense generates some new text at
context invariance, or changes the initial text changing the
contribution to context. The account of mutual the text and implied
sense influence, allows revealing heterogeneity of the text in view
of text and context streams. The define 4- divergence is possible as
4-vector of the text stream of \eqoref{diver} and also to calculate
antisymmetric tensor and dual to it tensor. These generally values
are not equal to zero. Thus, for preservation without dimension of
entered values, up to multiply the differential operator 4-
divergence on constant mean value of distance $\lambda_{c}$:
\begin{equation}
\begin{gathered}
\Sdiv{\alpha}\;I_{\alpha}\;\neq\;0\>,\mathtab
J^{\;\alpha\;\beta}\;=\;\left[\Sdiv{\alpha}\;I^{\beta}\;-\;\Sdiv{\beta}\;I^{\alpha}\right]
\;\neq\;0\>.\mathtab \tilde{J}^{\;\alpha\;\beta}\;=\;
\displaystyle{1\over
2}\;\varepsilon^{\;\alpha\;\beta\;\sigma\;\varrho}\;J_{\;\sigma\;\varrho}\neq\;0\>,
\end{gathered}
~\label{IJdef}
\end{equation}
The analogy to a 4-vector of the text stream is possible to enter
characteristics a 4-vector of context stream $B^{\;\alpha}$:
\begin{equation}
\begin{gathered}
\Sdiv{\alpha}\;B_{\;\alpha}\;\neq\;0\>,\mathtab
H^{\;\alpha\;\beta}\;=\;\left[\Sdiv{\alpha}\;B^{\;\beta}\;-\;\Sdiv{\beta}\;B^{\;\alpha}\right]
\;\neq\;0\>,\mathtab \tilde{H}^{\;\alpha\;\beta}\;=\;
\displaystyle{1\over
2}\;\varepsilon^{\;\alpha\;\beta\;\varrho\;\sigma}\;H_{\;\varrho\;\sigma}\neq\;0
\>.
\end{gathered}~\label{BHdef}
\end{equation}

Through the values certain thus can be expressed {\it a context
emotion}, as the sum various composed, describing interaction
between the text, implied sense and 4-vectors of text and implied
sense streams:
\begin{equation}
\begin{aligned}
\psi\;=\;
\psi_{0}\;+\;\psi_{I}\;+\;\psi_{B}+\;\psi_{J}\;+\;\psi_{H}\;+\;
\psi_{I B}\;+\;\psi_{I J}\;+\mathtab\psi_{B J}\;+\;\psi_{I
H}\;+\;\psi_{B H}\;+\;\psi_{J H}\>,
\end{aligned}~\label{psdef0}
\end{equation}
where $\psi_{0}$ -- is certain by \eqoref{Psdef}. ßThe obvious forms
of other composed are resulted below. The composed $\psi_{I}$
characterizes the contribution in {\it context emotion} from a
vector of the text stream in context, considering not only
self-action of the given vector, but also possible changes and
heterogeneity up to the second order inclusively:
\begin{equation}
\begin{aligned}
\IBdifF{\psi_{I}}{I}{I}{\nu}{\beta}
\IBdifS{I}{I}{\nu}{\beta}{\sigma} \>.\end{aligned}~\label{Ipsdef}
\end{equation}
The following composed $\psi_{B}$ characterizes the contribution in
{\it context emotion} from a vector of a context stream in a
context, considering not only self-action of the given vector, but
also its possible changes and heterogeneity up to the second order
inclusively:
\begin{equation}
\begin{aligned}
\IBdifF{\psi_{B}}{B}{B}{\nu}{\beta}
\IBdifS{B}{B}{\nu}{\beta}{\sigma}
{\varrho}\>,\end{aligned}~\label{Bpsdef}
\end{equation}
The composed $\psi_{J}$ characterizes the contribution in {\it
context emotion} from antisymmetric tensor of the text stream in a
context, at possible change and heterogeneity up to the second order
inclusively:
\begin{equation}
\begin{aligned}
\JHFdif{\psi_{J}}{J}{J}{\nu}{\beta}\mathtab
\JHSdif{J}{J}{\sigma}{\varrho}\mathtab
\JHTdif{J}{J}{\sigma}{\varrho}\>. \end{aligned}~\label{Jpsdef}
\end{equation}
The composed $\psi_{H}$ characterizes the contribution in {\it
context emotion} from antisymmetric tensor of context stream in a
context, at possible change and heterogeneity up to the second order
inclusively:
\begin{equation}
\begin{aligned}
\JHFdif{\psi_{H}}{H}{H}{\nu}{\beta}\mathtab
\JHSdif{H}{H}{\sigma}{\varrho}\mathtab
\JHTdif{H}{H}{\sigma}{\varrho} \>. \end{aligned}~\label{Hpsdef}
\end{equation}
The composed $\psi_{I B}$ considers possible contributions from
interaction of vectors of text stream and context stream:
\begin{equation}
\begin{aligned}
\IBCFdif{\psi_{I
B}}{I}{B}{\nu}~~~~~~~~~~~~~~~~~~~~~~~~~~~~~~~~~~~~~~~~~\mathtab
\IBCSdif{I}{B}{\beta}{\varrho}\mathtab
\IBCTdif{I}{B}{\varrho}\>.\end{aligned}~\label{IBpsdef}
\end{equation}
The composed $\psi_{I J}$ considers possible contributions from
interaction of text stream vectors and antisymmetric tensor of text
stream, including contributions from change of interaction character
between text stream vector, due to antisymmetric tensor of text
stream:
\begin{equation}
\begin{aligned}
\IBJHFdif{\psi_{I J}}{I}{I}{J}{\nu}{\beta}{1}~~~~~~~~~~~~~~\mathtab
\IBJHSdif{1}{J}{I}{\beta}{\sigma}\mathtab
\IBJHSdif{1}{\tilde{J}}{I}{\beta}{\sigma}\mathtab
\IBJHTdif{I}{J}{\nu}{\beta}{1}\mathtab
\IBJHFodif{I}{J}{\nu}{\beta}{\alpha}{1}\>.
\end{aligned}~\label{IJpsdef}
\end{equation}
The composed $\psi_{B J}$ considers possible contributions from
interaction of context stream vector and antisymmetric tensor of
text stream, including contributions from change of interaction
character between context stream vector, due to antisymmetric tensor
of text stream:
\begin{equation}
\begin{aligned}
\IBJHFdif{\psi_{B J}}{B}{B}{J}{\nu}{\beta}{2}~~~~~~~~~~~~\mathtab
\IBJHSdif{2}{J}{B}{\beta}{\sigma}\mathtab
\IBJHSdif{2}{\tilde{J}}{B}{\beta}{\sigma}\mathtab
\IBJHTdif{B}{J}{\nu}{\beta}{2}\mathtab
\IBJHFodif{B}{J}{\nu}{\beta}{\alpha}{2}\>,
\end{aligned}~\label{BJpsdef}
\end{equation}
The composed $\psi_{I H}$ considers possible contributions from
interaction of text stream vector and antisymmetric tensor of
context stream, including contributions from change of interaction
character between text stream vector, due to antisymmetric tensor of
context stream:
\begin{equation}
\begin{aligned}
\IBJHFdif{\psi_{I H}}{I}{I}{H}{\nu}{\beta}{1}~~~~~~~~~~~~~~\mathtab
\IBJHSdif{1}{H}{I}{\beta}{\sigma}\mathtab
\IBJHTdif{I}{H}{\nu}{\beta}{1}\mathtab
\IBJHFodif{I}{H}{\nu}{\beta}{\alpha}{1}
\>.\end{aligned}~\label{IHpsdef}
\end{equation}
The composed $\psi_{B H}$ considers possible contributions from
interaction of context stream vector and antisymmetric tensor of
context stream, including contributions from change of interaction
character between context stream vector, due to antisymmetric tensor
of context stream:
\begin{equation}
\begin{aligned}
\IBJHFdif{\psi_{B H}}{B}{B}{H}{\nu}{\beta}{2}~~~~~~~~~~~~\mathtab
\IBJHSdif{2}{H}{B}{\beta}{\sigma}\mathtab
\IBJHTdif{B}{H}{\nu}{\beta}{2}\mathtab
\IBJHFodif{B}{H}{\nu}{\beta}{\alpha}{2}\>.\end{aligned}~\label{BHpsdef}
\end{equation}
In the \eqsref{IJpsdef}{BHpsdef} are entered following multipliers
$k_1$ and $k_2$ facing composed. Their obvious form has been
presented through vectors of text stream and context stream:
\begin{equation}
\begin{aligned}
\FactorF{I}{\nu}{\alpha}{1}~~~~~~~~~~~~~~~\mathtab
\FactorS{I}{\alpha}{\beta}
\end{aligned}\>,~\label{K1f}
\end{equation}
\begin{equation}
\begin{aligned}
\FactorF{B}{\nu}{\alpha}{2}~~~~~~~~~~~~~~~~\mathtab
\FactorS{B}{\alpha}{\beta}\>.
\end{aligned}~\label{K2f}
\end{equation}
The composed $\psi_{J H}$ considers possible contributions from
interaction antisymmetric tensor of text stream and antisymmetric
tensor of context stream:
\begin{equation}
\begin{aligned}
\JHCAlldif{\psi_{J
H}}{J}{H}{\alpha}{\beta}{\sigma}{\varrho}{\delta}{\nu} \>.
\end{aligned}~\label{JHpsdef}
\end{equation}

Through the values certain thus has been possible to express {\it
implied sense}, as the sum of various contributions from the text
interaction with 4-vector of stream::
\begin{equation}
\gamma\;=\; \gamma_{\mu}\;+\;\gamma_{\psi}+\;\gamma_{\mu\;
\psi}\;+\;\gamma_{x}\>.~\label{rodef}
\end{equation}
The composed $\gamma_{\mu}$ characterizes the contribution from
changes and heterogeneity text emotions on which contributions to
implied sense of the perceived text depend:
\begin{equation}
\begin{gathered}
\gamma_{\mu}=\left(\Ddiv{2}\;\mu\right)\;+\;\left(\Sdivc{\alpha}\;\mu\right)
\left(\Sdiv{\alpha}\;\mu\right)\;+%\mathtab
\left(\Ddiv{2}\;\mu\right)\left(\Ddiv{2}\;\mu\right)\>.
\end{gathered}~\label{gmu}
\end{equation}
The composed $\gamma_{\psi}$ characterizes the contribution from
changes and heterogeneity context emotions on which contributions to
implied sense of perceived text to given context depend:
\begin{equation}
\begin{gathered}
\gamma_{\psi}=\left(\Ddiv{2}\;\psi\right)\;+\;
\left(\Sdivc{\alpha}\;\psi\right)\left(\Sdiv{\alpha}\;\psi\right)\;+%\mathtab
\left(\Ddiv{2}\;\psi\right)\left(\Ddiv{2}\;\psi\right)\>.
\end{gathered}~\label{gpsi}
\end{equation}
The composed $\gamma_{\mu\; \psi}$ characterizes the contribution
from interaction of changes and heterogeneity text and context
emotions on which contributions to implied sense of the perceived
text to the given context depend:
\begin{equation}
\begin{gathered}
\gamma_{\mu\;\psi}=\displaystyle{{1}\over{2}}\left[\left(\Sdivc{\alpha}\;\mu\right)
\left(\Sdiv{\alpha}\;\psi\right)\;+
\;\left(\Sdivc{\alpha}\;\psi\right)\left(\Sdiv{\alpha}\;\mu\right)\right]\;+\mathtab
\displaystyle{{1}\over{2}}\left[\left(\Ddiv{2}\;\mu\right)\left(\Ddiv{2}\;\psi\right)\;+\;
\left(\Ddiv{2}\;\psi\right)\left(\Ddiv{2}\;\mu\right)\right]\>.
\end{gathered}~\label{gpsimu}
\end{equation}
The composed $\gamma_{x}$ characterizes the contribution from
interaction of changes and heterogeneity text and a context
emotions, and also vectors of a stream of the text and a context, at
the account of possible contributions from antisymmetric tensors of
text streams and a context on which contributions to implied sense
of the perceived text to the given context expressed through the
moments calculated in the given point of space depend:
\begin{equation}
\begin{gathered}
\gamma_{x}\;=\;\displaystyle{{{1}\over{2}}\;
\left({{x_{\alpha}}\over{\lambda_{c}}}\right)}\left[\left(\Sdiv{\alpha}\;\mu\right)+
\left(\Sdiv{\alpha}\;\psi\right)+I^{\alpha}\;+\;B^{\alpha}\right]\;+\mathtab
\displaystyle{{1}\over{2}}\;\left[\left(\Sdivc{\alpha}\;\mu\right)+
\left(\Sdivc{\alpha}\;\psi\right)+I_{\alpha}+B_{\alpha}\right]
\;\displaystyle{\left({{x^{\alpha}}\over{\lambda_{c}}}\right)}\;+\mathtab
\displaystyle{\left({{x_{\alpha}}\over{\lambda_{c}}}\right)}\;\left[J^{\;\alpha,\;\beta}+
\tilde{J}^{\;\alpha,\;\beta}+H^{\;\alpha,\;\beta}+
\tilde{H}^{\;\alpha,\;\beta}\right]\;\displaystyle{\left({{x_{\beta}}\over{\lambda_{c}}}\right)}\>.
\end{gathered}~\label{gx}
\end{equation}
The apparently from the resulted parities for contributions to the
{\it GIE} the account of self-action leads, complex and
heterogeneity under contributions, structure of the {\it GIE}.
However, at the decision of specific targets, the part from this
composed can not be considered, proceeding from conditions of a
problem. So, for example at uniformity text perception components of
tensors, disappear both contributions to context and implied sense
of text and the {\it GIE} depends only on emotion of text.

\section{The description of free information transference}
The before to consider free information transference model have
necessary to consider that fact, that absolutely free information
transference does not represent interest as the information cannot
be characterized by the {\it GIE}, by virtue of absence of
information perception subject. The free information transference
has understood information perception for the other information
lack. The further has considered a method of processes description
at the free information transference.

The 4-DPEIS general postulates considered earlier and formal
structure {\it GIE} description, probably to consider the question
on free information transference in 4-DPEIS. The information
transference to 4-DPEIS has chance quantity, i.e. the information
position in the given 4-DPEIS point $x^{\alpha}_{b}$ can be found
with some probability of initial point $x^{\alpha}_{a}$ transition:
\begin{equation}
\Omega\left(x^{\;\alpha}_{b},\;x^{\;\alpha}_{a}\right)=\left|
K\left(x^{\;\alpha}_{b},\;x^{\;\alpha}_{a}\right)\right|^2=
K\left(x^{\;\alpha}_{b},\;x^{\;\alpha}_{a}\right)
K^{\ast}\left(x^{\;\alpha}_{b},\;x^{\;\alpha}_{a}\right)\>,~\label{Ver}
\end{equation}
where $K\left(b,\; a\right)$ -- is generally complex amplitude of
transition from a point $x^{\;\alpha}_{a}$ to a point
$x^{\;\alpha}_{b}$, $K^{\ast}\left(b,\; a\right)$ -- is complexly
interfaced amplitude. For transition amplitude calculation has used
method continual integrals of Feynman from action function
\cite{FeV00}. Thus, action for free transference information is
defined through a 4-vector of momentum:
\begin{equation}
S_{0}\left(x^{\;\alpha}_{b},\;x^{\;\alpha}_{a}\right)=
\int\limits_{\displaystyle{x^{\;\alpha}_{a}}}^{\displaystyle{x^{\;\alpha}_{b}}}\mathfrak{L}_{\;0}\;d\tau
=
-\int\limits_{\displaystyle{x^{\alpha}_{a}}}^{\displaystyle{x^{\;\alpha}_{b}}}
p^{\;\,\alpha}\;d x_{\,\alpha}\>.~\label{S0}
\end{equation}
The transition amplitude between points $x^{\;\alpha}_{a}$ and
$x^{\;\alpha}_{b}$ was possible to calculate as integral on all
possible paths from action function, where as a constant has been
used the certain value of information constant $\hbar_{c}$
\eqoref{hdef}:
\begin{equation}
K\left(x^{\;\alpha}_{b},\;x^{\;\alpha}_{a}\right)
=\displaystyle{{1}\over {N_{\infty}}}\int\limits_{\Gamma}
\exp\left[{\displaystyle{-{{i}\over{\hbar_{c}}}
S_{0}\left(x^{\;\alpha}_{b},\;x^{\;\alpha}_{a}\right)}}
\right]\prod\limits_{\displaystyle{x_{\;i}^{\;\alpha}}}
\displaystyle{{d^{\;4} p\left(x_{\;i}^{\;\alpha}\right)\;d^{\;4}
x\left(x_{\;i}^{\;\alpha}\right)}\over{8\; \pi^4}}\>,~\label{Kdef}
\end{equation}
where $\Gamma$ -- is a of phase space volume in which there is the
information transference on all possible paths; $x_{\;i}^{\;\alpha}$
-- is carried out $i$-partition number of information position
4-vector in 4-DPEIS, partition trajectory which moving the
information from initial point $x^{\;\alpha}_{a}$ in finite point
$x^{\alpha}_{b}$, it essentially depends on partition paths;
$d^{\;4} p\left(x_{i}^{\;\alpha}\right)$ è $d^{\;4}
x\left(x_{i}^{\;\alpha}\right)$ -- is corresponding $i$-partition
number to that splitting of phase 4-volume paths elements in
momentum and coordinate spaces; $N_{\infty}$ -- is normalizing
constant which can be defined from normalizing condition on unit of
probability in all 4-DPEIS:
\begin{equation}
\begin{gathered}
K\left(\Gamma_{\infty}\right)\;
=\;\displaystyle{\int\limits_{\Gamma_{\infty}}
\exp\left[{\displaystyle{-{{i}\over{\hbar_{c}}}\;
S_{0}\left(x^{\;\alpha}_{b},x^{\;\alpha}_{a}\right)}}
\right]\prod\limits_{\displaystyle{x_{i}^{\;\alpha}}}\;
\displaystyle{{d^{\;4} p\left(x_{i}^{\;\alpha}\right)\;d^{\;4}
x\left(x_{i}^{\;\alpha}\right)}\over{8\; \pi^4}}}\>,\mathtab
\left|\displaystyle{{1}\over
{N_{\infty}}}\;K\left(\Gamma_{\infty}\right)\right|^{2}\;=\;
\displaystyle{{1}\over {N_{\infty}
N^{\ast}_{\infty}}}\;K\left(\Gamma_{\infty}\right)K^{\ast}\left(\Gamma_{\infty}\right)\;
=\;1\>,
\end{gathered}~\label{Normdef}
\end{equation}
where $\Gamma_{\infty}$ -- is all 4-volume in 4-DPEIS of the given
information extends on all possible paths. The obvious form of paths
integral have calculated between two points $x^{\;\alpha}_{a}$ and
$x^{\;\alpha}_{b}$ of \eqoref{Kdef} which defines amplitude of
information transition, and then obvious form of normalizing
constant have calculate for free information transference frame of
\eqoref{Normdef}. The calculation of paths integral is spent by
partition each path into such sites, where
$p^{\;\alpha}\left(x_{\;i}^{\;\alpha}\right)$ -- is constant
everyone $i$-partition number,
$x^{\;\alpha}\left(x_{\;i}^{\;\alpha}\right)$ -- is varies linearly,
the $M$ parts number for each of paths will depend on away
partition, nevertheless, in a limit at $M\;\rightarrow\;\infty$ the
following obvious forms is received for continual integral:
\begin{equation}
K\left(b,\; a\right) =\displaystyle{{1}\over
{N_{\infty}}}\int\limits_{\Gamma}
\exp\left[{\displaystyle{-{{i}\over{\hbar_{c}}}\;
p_{\;\alpha}\left(x^{\alpha}_{b}\;-\;x^{\alpha}_{a}\right)}}\right]\;
\displaystyle{{d^{\;4} p}\over{8\; \pi^4}}\>.~\label{Kint}
\end{equation}
The integral of \eqoref{Kint} will be possible calculate, knowing an
obvious form of 4-vector of momentum $p^{\;\alpha}$. For calculation
of mean value is necessary know in obvious form 4-density of the
information transition probability from initial trajectory in final
point. For this purpose have considered probability of such
transition \eqoref{Ver} and have find density of probability in
coordinate space:
\begin{equation}
\begin{aligned}
\rho\left(x^{\;\alpha}_{b},\;x^{\;\alpha}_{a}\right)\;=\;
\displaystyle{{d\;\Omega\left(x^{\;\alpha}_{b},\;x^{\;\alpha}_{a}\right)}\over{d^{\;4}
x}}\;=~~~~~~~~~~~~~~~~~~~~~~~~~~~~~~~~~~~~~~~~~~~~~~~\mathtab
\displaystyle{{d\;K\left(x^{\;\alpha}_{b},\;x^{\;\alpha}_{a}\right)}\over{d^{\;4}
x}}K^{\ast}\left(x^{\;\alpha}_{b},\;x^{\;\alpha}_{a}\right)+
K\left(x^{\;\alpha}_{b},\;x^{\;\alpha}_{a}\right)
\displaystyle{{d\;K^{\ast}\left(x^{\;\alpha}_{b},\;x^{\;\alpha}_{a}\right)}\over{d^{\;4}
x}}\>.
\end{aligned}~\label{Rodef}
\end{equation}
The received definition of probability density can be used at
calculation of mean from observable values: 4-vectors of information
displacement, a 4-vector of momentum, etc. Also the given expression
\eqoref{Rodef} can be used for updating model parameters in
requirements view \theoremref{Ther2}, i.e. at calculation of mean
displacement in 4-DPEIS:
\begin{equation}
\langle \left(x^{\;\alpha}_{b}\;-\;x^{\;\alpha}_{a}\right)\rangle =
\int\limits_{\Gamma}
\left(x^{\;\alpha}_{b}\;-\;x^{\;\alpha}_{a}\right)
\rho\left(x^{\;\alpha}_{b},\;x^{\;\alpha}_{a}\right)\;{d^{\;4} x}\>.
~\label{Delxdef}
\end{equation}

The method offered in the given paragraph can be used for
characteristics renewal of free information apprehended by the
subject of information perception: knowledge of obvious two
information characteristics forms, such as the {\it GIE} of
\eqoref{qmgpdef}, tensor of text of \eqoref{text} and  tensor of
text perception of \eqoref{imput} allows to calculate all three
characteristics and to calculate depending from them observable.

In the given section the basic moments of the free information
transference description in 4-DPEIS and calculations are considered
the characteristics related, and also the normalizing constant which
can be used in the further is calculated. In the further shall
consider the problem on information transference at interference on
it of other information.

\section{The description of information interaction and transference}
The considering 4-DPEIS calculating information interaction
characteristics is necessary to consider, that interference of
information occurs concerning one subject of perception of the
information, in frame of if there are some subjects of perception of
the information that interference of information occurs due to
information interchange between these subjects. The consideration of
information interference leads to that concerning one subject, at
information difference perception could be considered of information
concerning of one information perception tensor of \eqoref{imput}.
The observable values for information interaction have chance
quantity, therefore for calculations is necessary for us to know
processes probability and probability density.

For calculation of system transition probability from some initial
point $x^{\alpha}_{a}$ in final point $x^{\alpha}_{b}$ èin 4-DPEIS
space is come return to consideration Lagrangian of two interaction
information system of \eqoref{LagB}. The Lagrangian of two
interaction information system was entered the effective 4-vector of
momentum, taking into account of \eqoref{LagBD}:
\begin{equation}
P^{\;\alpha}\;=\; \;Q_{c}\;\nu_{c}\;\left( n_{+}\;q_{+}
\left[f\;\beta_{\;\alpha\;+}+{{q_{1 2}} \over {\left[s\left(
x^{\;\alpha}\right)\right]^{\,2}}}\;
a^{\;\alpha}\right]\right)\>.~\label{MomInt}
\end{equation}
For total 4-velocity of system is entered designation still in
addition of \eqoref{LagBD}:
\begin{equation}
v^{\;\alpha}\;=\; \nu_{c}\;\beta_{\;\alpha\;+}\>.~\label{VelInt}
\end{equation}
The result was accepted following form for Lagrangian of two
interaction information system:
\begin{equation}
\displaystyle{\mathfrak{L}_{\;int} = P^{\;\alpha}\;
v_{\;\alpha}}\>.~\label{LagBÑ}
\end{equation}
The system transition probability of two information interaction, as
well as in frame of free transference, is defined through complex
amplitude:
\begin{equation}
\Omega_{\;int}\left(x^{\;\alpha}_{b},\;x^{\;\alpha}_{a}\right)=
K_{\;int}\left(x^{\;\alpha}_{b},\;x^{\;\alpha}_{a}\right)
K_{\;int}^{\ast}\left(x^{\;\alpha}_{b},\;x^{\;\alpha}_{a}\right)\>,~\label{VerIn}
\end{equation}
where $K_{int}\left(b,\; a\right)$ -- is a complex amplitude of two
information system transition from point $x^{\;\alpha}_{a}$ in point
$x^{\;\alpha}_{b}$, $K^{\ast}\left(b,\; a\right)$ -- is complex
conjugate amplitude. For calculation of transition amplitude is used
continual integrals method from action function, offered Feynman
\cite{FeV00}. Thus action for free transference information is
defined through 4-vector of momentum:
\begin{equation}
S_{\;int}\left(x^{\;\alpha}_{b},\;x^{\;\alpha}_{a}\right)=
\int\limits_{\displaystyle{x^{\;\alpha}_{a}}}^{\displaystyle{x^{\;\alpha}_{b}}}\mathfrak{L}_{\;int}\;d\tau
=
-\int\limits_{\displaystyle{x^{\alpha}_{a}}}^{\displaystyle{x^{\;\alpha}_{b}}}
P^{\;\,\alpha}\;d x_{\,\alpha}\>.~\label{Sintn}
\end{equation}
The two information transition amplitude between points
$x^{\;\alpha}_{a}$ and  $x^{\;\alpha}_{b}$ is possible to calculate
as path integral from action function, where as a constant is used
earlier the information constant certain value $\hbar_{c}$ of
\eqoref{hdef}:
\begin{equation}
K\left(x^{\;\alpha}_{b},\;x^{\;\alpha}_{a}\right)
=\displaystyle{{1}\over {N_{\infty}}}\int\limits_{\Gamma}
\exp\left[{\displaystyle{-{{i}\over{\hbar_{c}}}
S_{int}\left(x^{\;\alpha}_{b},x^{\;\alpha}_{a}\right)}}\right]
\prod\limits_{\displaystyle{x_{\;i}^{\;\alpha}}}
\displaystyle{{d^{\;4} P\left(x_{\;i}^{\;\alpha}\right)d^{\;4}
x\left(x_{\;i}^{\;\alpha}\right)}\over{8\; \pi^4}}
\>,~\label{Kintdef}
\end{equation}
where $\Gamma$ -- is phase space volume is information transference
on all possible paths; $x_{\;i}^{\;\alpha}$ -- is $i$-partition
number of information position 4-vector in the 4-DPEIS, describing
path on which moving information is carried out from initial point
$x^{\;\alpha}_{a}$ to finite point $x^{\alpha}_{b}$, essentially is
depends approach of partition paths, $d^{\;4}
P\left(x_{i}^{\;\alpha}\right)$ and $d^{\;4}
x\left(x_{i}^{\;\alpha}\right)$ -- corresponding $i$-partition
number of phase 4-volume parting elements in momentum and coordinate
spaces; $N_{\infty}$ -- in frame of a free information system, i.e.
systems without interference, normalizing constant of
\eqoref{Normdef} which have defined from normality condition on
probability unit in all 4-DPEIS.

Let's obvious form of path integral calculate which defines two
information transition amplitude between two points
$x^{\;\alpha}_{a}$ and $x^{\;\alpha}_{b}$ of \eqoref{Kintdef}, and
normalizing constant of \eqoref{Normdef} for free information
transference frame have calculate obvious form. The calculation of
paths integral is spent by parting each path into such part, i.e.
also as well as in frame of free information transference, where
$P^{\;\alpha}\left(x_{\;i}^{\;\alpha}\right)$ -- is constant
everyone $i$-partition number into such part,
$x^{\;\alpha}\left(x_{\;i}^{\;\alpha}\right)$ -- varies linearly,
the $M$ parts number for each of paths will depend on away
partition, for each of paths will depend on away partition,
nevertheless, in a limit at $M\;\rightarrow\;\infty$ the following
obvious forms is received for continual integral:
\begin{equation}
K\left(b,\; a\right) =\displaystyle{{1}\over
{N_{\infty}}}\int\limits_{\Gamma}
\exp\left[{\displaystyle{-{{i}\over{\hbar_{c}}}\;
P_{\alpha}\left(x^{\alpha}_{b}\;-\;x^{\alpha}_{a}\right)}}\right]\;
\displaystyle{{d^{\;4} P}\over{8\; \pi^4}}\>.~\label{KintP}
\end{equation}
The integral of \eqoref{KintP} calculating is necessary solve for
obvious form of 4-vector of effective momentum $P^{\;\alpha}$.
However is possible pass from 4-volume integration element of
effective momentums in phase space, get 4-vector of momentum and
4-vector of interference potential obvious forms.

For calculation of mean values are necessary us known probability
4-density in obvious form of information transition from initial
path in finite point. For this purpose has considered probability of
\eqoref{Ver} such transition and found probability 4-density in
coordinate space:
\begin{equation}
\begin{aligned}
\rho\left(x^{\;\alpha}_{b},\;x^{\;\alpha}_{a}\right)\;=\;
\displaystyle{{d\;\Omega_{\;int}\left(x^{\;\alpha}_{b},\;x^{\;\alpha}_{a}\right)}\over{d^{\;4}
x}}\;=~~~~~~~~~~~~~~~~~~~~~~~~~~~~~~~~~~~~~~~~~~\mathtab
\displaystyle{{d\;K\left(x^{\;\alpha}_{b},\;x^{\;\alpha}_{a}\right)}\over{d^{\;4}
x}}K^{\ast}\left(x^{\;\alpha}_{b},\;x^{\;\alpha}_{a}\right)+
K\left(x^{\;\alpha}_{b},\;x^{\;\alpha}_{a}\right)
\displaystyle{{d\;K^{\ast}\left(x^{\;\alpha}_{b},\;x^{\;\alpha}_{a}\right)}\over{d^{\;4}
x}}\>.
\end{aligned}~\label{Rointdef}
\end{equation}
The received definition of probability 4-density can be used at
calculation of averages from observable values: 4-vectors of
information displacement, 4-vector of momentum, etc. Also the given
of \eqoref{Rointdef} can be used for updating parameters of model in
requirements view \theoremref{Ther2}.

The paragraph conclusions will lead classification of arising
problems by information interference, under processing condition of
these information (it is corresponding texts tensors of
\eqoref{text}) by means of one information perception tensor of
\eqoref{imput}, i.e. the perception of the information is made by
one subject
\cite{Aul05,HoO05,HoS05,She48,She49,She50,She51,She49MP,Kol91,Haz98,AsZ07,
JoV07,HiM07,Mas07,ArL96,MaC97,HaC98,Kan02,BaV04,Kan04,Koc04,MiB03,Gim06,
BoR96,HeB01BCFGM,RiB01,HeB01,BoR05,MoC07,Tat89,Koz91,LoB05,FaB06,Bur06,Per07,Zie07,Mik07}:
\begin{enumerate}
\item The problem of information transference under other information interaction
is primary goal about information interference, when by known
obvious form of tensor of text perception and tensor of texts
participating in interference, the {\it GIE} obvious form of
\eqoref{qmgpdef}. The offered method will be calculating probability
of information transference and probability 4-density and also will
be possible to calculate all observable parameters, when knowing
velocity of information transference in 4-DPEIS.
\item The return problem of information transference under other information interaction:
knowing {\it GIE} obvious form observable parameters of information
interference, and also obvious form of tensor of text perception and
one of tensor of text and one only of two tensor of texts
participating in interference will be possible to restore
accordingly initial tensor of text or tensor of text perception. The
same problem will be carried problem of generation new tensor of
text under interaction of apprehended text.
\item The problem of formation information perception tensor
(the problem of the subject training): unlike a problem of renewal
information perception tensor could be used tensor of texts with in
advance known {\it GIE} and on them correct value {\it GIE} for each
of giving separately texts will be calculated the tensor of text
perception.
\item The problem of information dispersion will be problem existence
of information barrier to information perception.
\end{enumerate}

\section{Results and conclusions}

So, in this article the direction of construction of information
space is considered as 4-DPEIS where time is the value equivalent to
time in physical system with which the information is related. Thus
we can make following statements:
\begin{enumerate}
\item criteria of existence {\it LIV} of \eqoref{Cvel} are found and value determined is presented
\theoremref{Ther1}
\item On the basis of existence {\it LIV} can enter 4-DPEIS,
on which the Poincares group of transformations  determined for {\it
LIV} of \eqref {Cvel} is realized
\item criteria of the information displacement in information space \theoremref{Ther2}
which establish communication between mean time, are determined by
mean velocity and the information displacement .
\item For the description of information dynamics and interaction is entered the
generalizing parameter --- {\it GIE}, the basic components, giving
the contribution to this parameter, and the equations relating this
parameter with parameters of model are considered.
\item the description of free information transference and
interference of two and more information in information space is
considered. The basic classification of problems is resulted.
\end{enumerate}

The considered direction of modelling of information space can be
applied without the generality limitations to the description of any
real systems for which studying information making system is
important and thus the account of processes of the various nature,
for example, physical and sociological processes is required. The
given direction of of information space modelling  allows to
consider, on the one hand, objective parameters of the information,
on the other hand, allows to consider within the limits of the
approach subjective features of perception of the information. The
important question at modelling information space, and also the text
and the tensor of text perceptions is a correct comparison of the
phenomena and processes proceeding in physical (chemical,
biological, social, etc.) system with three-dimensional coordinates
of information space and components the tensor of text and tensor of
text perception. As a whole, apparently to the author, the given
relativistic direction of the modelling description of information
space and information systems has greater prospects from the point
of view of the description of information processes in one language
and without dependence from type of considered system or especially
their combinations.

% ****************************************************

\ack {The author grateful to A.V.Gorokhov., A.F.Krutov and A.A.
Mironov for attention to the given this article and for numerous and
fruitful discussions.}

\end{document}